\documentclass[aps,prd,amssymb,nofootinbib,twocolumn,floatfix]{revtex4}
\usepackage{amssymb}
\usepackage{amsmath}
\usepackage{latexsym}
\usepackage{bm}

\newcommand{\half}{{\scriptstyle\frac{1}{2}}}

\newcommand{\DeltaI}{\Delta^{\kern -0.2em(1)}}
\newcommand{\DeltaII}{\Delta^{\kern -0.2em(2)}}
\newcommand{\DeltaN}{\Delta^{\kern -0.2em(n)}}
\newcommand{\deltaI}{\delta^{(1)}}
\newcommand{\deltaII}{\delta^{(2)}}
\newcommand{\deltaN}{\delta^{(n)}}

\newcommand{\deltaIIRR}{\delta^{(2)}_R}

\newcommand{\xiII}{\xi^{(2)}{}}
\newcommand{\xiI}{\xi^{(1)}{}}

\newcommand{\bxiI}{{\bm\xi}^{(1)}{}}
\newcommand{\bxiII}{{\bm\xi}^{(2)}{}}
\newcommand{\bxi}{{\bm\xi}}
\newcommand{\bXi}{{\bm\Xi}}

\usepackage[normalem]{ulem}
\newcommand{\stkout}[1]{\ifmmode\text{\sout{\ensuremath{#1}}}\else\sout{#1}\fi}
\usepackage{hyperref}
\hypersetup{
  colorlinks=true,        
  linkcolor=cyan,         
  citecolor=cyan,         
}

\newcommand{\chial}{\chi_\alpha}

\newcommand{\be}{\begin{equation}}
\newcommand{\ee}{\end{equation}}
\newcommand{\bea}{\begin{eqnarray}}
\newcommand{\eea}{\end{eqnarray}}
\newcommand{\beaa}{\begin{eqnarray*}}
\newcommand{\eeaa}{\end{eqnarray*}}
\newcommand{\pa}{\partial}
\newcommand{\na}{\nabla}
\newcommand{\Lie}{\mbox{\pounds}}

\newcommand{\bsube}{\begin{subequations}}
\newcommand{\esube}{\end{subequations}}

\newcommand\inner[2]{{\bm\langle}{#1}\bm|{#2}\bm\rangle}

\newcommand{\compcent}[1]{\vcenter{\hbox{$#1\circ$}}}
\newcommand{\comp}{\mathbin{\mathchoice
  {\compcent\scriptstyle}{\compcent\scriptstyle}
  {\compcent\scriptscriptstyle}{\compcent\scriptscriptstyle}}}
\DeclareMathAlphabet{\mathpzc}{OT1}{pzc}{m}{it} 
\begin{document}

\title{Limits on Magnetic Field Amplification from the r-Mode Instability}

\author{John L. Friedman${}^1$,  Lee Lindblom${}^{2}$,
Luciano Rezzolla${}^{3,4}$, and Andrey I. Chugunov${}^5$}

\affiliation{${}^1$Leonard Parker Center for Gravitation, Cosmology and Astrophysics,
University of Wisconsin-Milwaukee, P.O. Box 413,
Milwaukee, Wisconsin 53201, USA}
\email{friedman@uwm.edu}

\affiliation{${}^2$Center for Astrophysics and Space Sciences, University of California at San Diego, La Jolla, CA 92093, USA}
\email{llindblom@ucsd.edu}

\affiliation{${}^3$Institute for Theoretical Physics,
  Max-von-Laue-Stra{\ss}e 1, 60438 Frankfurt, Germany}
\email{rezzolla@itp.uni-frankfurt.de}

\affiliation{${}^4$Frankfurt Institute for Advanced Studies, Ruth-Moufang-Str. 1,
60438 Frankfurt, Germany}

\affiliation{${}^5$Ioffe Institute, Polytekhnicheskaya 26, 194021 St.-Petersburg, Russia}
\email{andr.astro@mail.ioffe.ru}

\date{\today}

\begin{abstract}
At second order in perturbation theory, the unstable $r$-mode of a
rotating star includes growing differential rotation whose form and
growth rate are determined by gravitational-radiation reaction. With no
magnetic field, the angular velocity of a fluid element grows
exponentially until the mode reaches its nonlinear saturation amplitude
and remains nonzero after saturation. With a background magnetic field,
the differential rotation winds up and amplifies the field, and previous
work where large mode amplitudes were considered \cite{Rezzolla00},
suggests that the amplification may damp out the instability. A
background magnetic field, however, turns the saturated time-independent
perturbations corresponding to adding differential rotation into
perturbations whose characteristic frequencies are of order the Alfv\'en
frequency. As found in previous studies, we argue that
  magnetic-field growth is sharply limited by the saturation amplitude of
  an unstable mode. In contrast to previous work, however, we show that
  if the amplitude is small, i.e.,~$\lesssim 10^{-4}$, then the limit on
  the magnetic-field growth is stringent enough to prevent the loss of
  energy to the magnetic field from damping or significantly altering an
  unstable $r$-mode in nascent neutron stars with normal interiors and in
  cold stars whose interiors are type II superconductors.
 We show this result first for a toy model, and we then
obtain an analogous upper limit on magnetic-field growth using a more
realistic model of a rotating neutron star.
Our analysis depends on the assumption that there are no
marginally unstable perturbations, and this may not hold when
differential rotation leads to a magnetorotational instability.

\end{abstract}

\maketitle

\section{Introduction}
\label{s:intro}
\noindent

Gravitational radiation drives an instability in the $r$-modes of
rotating relativistic stars \cite{A97,FM98} whose growth
time~\cite{Lindblom98b} may be short enough to limit the angular velocity
of old accreting neutron stars and may contribute to the spin-down of
nascent neutron stars (see
\cite{Lindblom98b,fsbook,Bondarescu07,Bondarescu09,h15,gck16} for reviews
and references).  At second order in perturbation theory, the unstable
mode includes exponentially growing differential rotation
\cite{Rezzolla00,Rezzolla01b,Rezzolla01c,LevinUshormirsky01,Sa2004,fll15},
whose form with no magnetic field was recently obtained by Friedman,
Lindblom and Lockitch \cite{fll15} (henceforth {\it Paper I}). Past work
that considered $r$-modes saturated at large amplitudes in newly born and
highly magnetized neutron stars has suggested that the resulting magnetic
field windup could damp out or significantly alter the instability
\cite{Spruit99,Rezzolla00,Rezzolla01b,Rezzolla01c,Cuofano2010,Cuofano_etal12,CZW15}.
The present paper, however, which considers smaller saturation
amplitudes, finds restrictions on the growth of differential rotation
that appear stringent enough to exclude significant damping of the
instability by magnetic fields in old neutron stars spun up by accretion
and in nascent, rapidly rotating
stars.  For the stable $r$-mode, with no radiation reaction, the secular
drift is pure gauge \cite{Chugunov2015}: It can be removed by
adding a second-order time-independent perturbation that adds
differential rotation to the unperturbed equilibrium star.

The growth of an unstable mode is limited by nonlinear saturation -- that
is, by loss of energy to other modes at a rate equal to the growth rate
of the unstable mode. In their studies of magnetic field windup by an
unstable $r$-mode in nascent neutron stars, Rezzolla {\it et al.}
\cite{Rezzolla00,Rezzolla01b,Rezzolla01c,Cuofano2010} use a saturation
amplitude $\alpha_{\rm sat}$ of order $10^{-1}$ or larger, as these were
the typical values estimated to be relevant in newly born neutron stars
\cite{Owen1998}. Subsequent work in the context of second-order
perturbation theory, however, finds an amplitude smaller than $10^{-4}$
\cite{Arras_etal03,BTW04,BTW05,Bondarescu07,BW13}, and recent papers
argue for still smaller limits based on observations of low-mass X-ray
binaries and millisecond pulsars \cite{gck16,ms13}.
Although a small saturation amplitude in itself sharply limits the effect
of magnetic-field windup on the $r$-mode instability of young stars,
Cuofano {\it et al.} \cite{Cuofano2010,Cuofano_etal12} find a substantial
effect on $r$-mode evolution in old accreting neutron stars.  They use
the formalism developed by Rezzolla {\it et al.}
\cite{Rezzolla00,Rezzolla01b,Rezzolla01c}. They do not include nonlinear
couplings, but the amplitude in their simulations remains below
$10^{-4}$. What these studies do not include is the back-reaction of
magnetic field windup on the second-order perturbation associated with
differential rotation, and that is the focus of the present work.

For a stationary star with no magnetic field and no viscosity,
adding differential rotation is a time-independent perturbation: It
simply changes a uniformly rotating equilibrium to a neighboring
equilibrium with a slightly different rotation law. Still in the
  absence of viscosity, but with a background magnetic field, however, a
perturbation that adds differential rotation is a sum of axisymmetric
modes with nonzero frequencies, modes restored by the magnetic Lorentz
force -- by the tension of stretched field lines. The periods of these
modes are of order the Alfv\'en time $t_A$, which is essentially the
time over which a perturbation in the magnetic field travels across a
reference lengthscale in a plasma, which we take here to be the radius
$R$ of the star.

At second order in perturbation theory, differential rotation of an
unstable star with negligible magnetic field is driven by a second-order
radiation-reaction force together with quadratic terms in the perturbed
magnetohydrodynamics (MHD)-Euler equation (terms quadratic in the perturbed variables of
the first-order $r$-mode). Before saturation, the effective driving force
grows exponentially over a gravitational radiation-reaction timescale
$\tau_{GR}$, driving an exponentially growing differential rotation.
After saturation, the driving force is constant, but the differential
rotation maintains a power-law growth in time.

With a magnetic field large enough that $t_A \lesssim \tau_{GR}$ and a
sufficiently small saturation amplitude, the picture is sharply
altered. Now the driving force acts on a set of axisymmetric modes with
frequencies of order $\omega_A =2\pi/t_A$. Before saturation, the
amplitudes of these modes again grow exponentially. But after saturation,
each of the modes that comprise the differential rotation is effectively
an oscillator acted on by a constant force: Its amplitude is the sum of
its amplitude at saturation and a solution with harmonic time
dependence. The combination of the small-saturation amplitude of the
first-order $r$-mode and the fact that the growth of second-order
differential rotation stops shortly after saturation, leads to a
stringent constraint on differential rotation (on the secular drift of a
fluid element) and hence on magnetic-field windup. We find that the
increase in the magnetic field prior to saturation is smaller than the
value needed to damp the unstable $r$-mode by a factor of order
$\alpha$; equivalently, the rate at which the magnetic field's energy
drains energy from the $r$-mode is smaller by a factor of order
$\alpha^2$ than the rate at which the radiation-reaction force drives
the unstable mode\footnote{In Ref.~\cite{Chugunov2015}, Chugunov notes
an analogous relation for the stable $r$-mode if one
assumes that the arbitrarily chosen initial differential rotation is
of order $\alpha^2$. Here, for the unstable $r$-mode, the induced
differential rotation is necessarily of order $\alpha^2$, but one
needs an additional constraint (Eq.~\eqref{e:xi_bound1} below) to
keep the secular exponential growth of the magnetic field below its critical value.}.
When $\alpha \sim \mathcal{O}(1)$, as assumed in the initial
investigations of the instability \cite{Owen1998} and in
Refs.~\cite{Rezzolla00, Rezzolla01b, Rezzolla01c}, this difference is
small, but the situation changes considerably if $\alpha \sim 10^{-4}$,
as in the present study.

The major results of this paper can be summarized as follows.  In
  Sec.~\ref{s:magnitudes} we qualitatively describe the fundamental
  physical processes that contribute to this problem: the timescales
  associated with the r-mode fluid oscillations, the timescales
  associated with magnetic field processes, and the timescale on which
  gravitational radiation drives an r-mode toward instability in
  neutron stars.  We summarize in Sec.~\ref{s:magnitudes} previously published
  estimates of the magnetic field strength needed to suppress the
  growth of the gravitational radiation-driven r-mode instability in
  neutron stars. The section ends with an outline of the argument 
  that gives our main result. 

In Sec.~\ref{s:toy}, we introduce a modified version of a toy model due
to Shapiro \cite{shapiro00} that illustrates the main features we have
just discussed. In Shapiro's model a cylinder of uniform-density fluid
with an initial magnetic field and initial differential rotation has a
time evolution given by the MHD-Euler system in the
ideal-magnetohydrodynamics (MHD) limit (i.e., in a plasma with
infinite conductivity). We add to the system a forcing term that
mimics the second-order axisymmetric radiation reaction force. Although
the system is non-perturbative, the fluid displacement and magnetic field
satisfy linear equations and can be written as a superposition of normal
modes. We find an analytic solution for its evolution and use it to
obtain a first estimate of the maximum angular displacement and magnetic
field of the $r$-mode.

In Sec.~\ref{s:perturbation}, we develop the formalism governing the
equilibrium and first- and second-order perturbations of a rotating star
with a background magnetic field, in an ideal-MHD
(MHD) framework with
radiation-reaction.  We express perturbations in terms of a Lagrangian
displacement and obtain the second-order MHD-Euler equation.  In contrast
to the toy model, the equation involves terms with first as well as
second time derivatives, and we need a formalism developed by Dyson and
Schutz \cite{DS79}, based on a conserved symplectic product \cite{fs78},
to express the amplitude of each mode in terms of the effective driving
force.

In Sec.~\ref{s:estimates}, we obtain estimates of the maximum angular
displacement of a fluid element and on the corresponding magnetic-field
amplification for the second-order unstable $r$-mode itself. We assume
that the perturbations are governed by a barotropic equation of state,
that axisymmetric perturbations of the equilibrium star conserving
angular momentum and baryon number are strictly stable, and that such
axisymmetric perturbations can be written as a sum of discrete,
nondegenerate modes. A brief discussion in
Sec. \ref{sec:discussion} summarizes our conclusions and considers
implications of relaxed assumptions.

We relegate to appendices details of the Lagrangian perturbation theory
and of the formalism that obtains the amplitude of fluid modes in
terms of a driving force.

\section{Underlying magnitudes}
\label{s:magnitudes}

\subsection{A problem with four timescales}
\label{s:timescales}

Four timescales are involved in this problem. In order of
increasing size
are: (1) the rotation period $2\pi/\Omega$ of the star; (2) the
oscillation period $T_{\rm mode} = 2\pi/\omega$ of an $r$-mode; and
(3) the $r$-mode growth time $\tau_{\rm mode}$. Timescale (4),
the Alfv\'en time $t_A$, may be larger or smaller than
 $\tau_{\rm mode}$, depending on the magnitude of the initial magnetic
field and on whether the neutron star's interior is superconducting.

The $r$-mode frequency $\omega$ is proportional to the star's
angular velocity, having for slowly rotating Newtonian stars the form
\be
  \omega = -\frac{(\ell-1)(\ell +2)}{\ell+1}\Omega\,,
\ee
for a mode associated with the $\ell=m$ angular harmonic. The critical
rotational frequency above which the $\ell=m =2$ $r$-mode is unstable
depends sensitively on temperature, but is likely to be above
$f=\Omega/(2\pi)\simeq 500$ Hz, and the corresponding periods of
rotation and oscillation are then of order 1-2 ms.

We define the Alfv\'en velocity $v_A$ for a normal plasma by
\be
   v_A = \sqrt{\frac{B^2}{4\pi\rho} }\ .
\ee
Using the radius $R$ of the star as a characteristic wavelength gives
the corresponding Alfv\'en angular frequency
\be
  \omega_A = 2\pi v_A/R = \frac BR\sqrt{\frac\pi\rho},
\label{e:omega_A}\ee
where $\rho$ is an average rest-mass density
\cite{bk74,tv86,romani90}. In old and accreting neutron stars,
such as those in X-ray binaries,
the corresponding magnetic fields inferred from observation
are in the
range $10^8$ - $10^9$ G. The interior poloidal and toroidal fields may
be higher, with the exterior poloidal field partly suppressed
by the accreting material \cite{bk74,tv86,romani90}, and the
relative size of the poloidal/toroidal magnetic-field components
remains an open question \cite{Ciolfi2013}. Using the inferred values
and typical sizes and densities for neutron stars, the typical Alfve\'n
timescale for a normal plasma is
\be
 t_A \sim {R\over v_A} =  7\times 10^4
  \ R_6\,B_9^{-1} \,\sqrt{\rho_{14.6}}\ \ {\rm s} ,
\label{e:t_Anormal}\ee
where $B$ is an average magnetic field intensity, and the
subscripts refer to Gaussian-cgs units, e.g.,
\mbox{$R_6 := R/(10^6$ cm)}. This timescale is considerably
shorter if the neutron star interior is a type II superconductor,
in which case the magnetic field is confined to flux tubes carrying
fields of order $H_c \gtrsim 10^{15}$ G and the Alfv\'en time is of order
\be
   t_{A, SC} \sim R \sqrt{ \frac{4\pi\rho}{B H_c}} \sim 70
\ R_6\sqrt{ \frac{\rho_{14.6}}{B_{9}\, H_{c,15}}} \ {\rm s}\,.
\label{e:t_A}\ee
Nascent neutron stars have normal interiors and observed magnetic fields
that range from $10^{12}$ to $10^{15}$ G.

Finally, the growth time $\tau_{\rm mode}$ of the $r$-mode instability is set
by a competition between gravitational radiation reaction and local
dissipation; the dominant contribution to local dissipation may be shear
viscosity for a normal interior or at the core-crust interface, or mutual
friction for a dominantly superfluid interior. In the absence of
viscosity, the growth time of the instability is the gravitational
radiation-reaction timescale, given for an equation of state with 
average polytropic index of order $0.5$ by \cite{Lindblom98b,ak01}
\be
\tau_{GR}\sim 2\times 10^3\, f_{500}^{-6}\, \frac{1.4M_\odot}{M}R_6^{-4}\ {\rm s}\,,
\label{e:tau_GR}\ee
where, adopting 500 Hz as a fiducial rotational frequency, we write
$f_{500} := f/500$ Hz.
Below a critical frequency, viscosity damps the instability. An 
accreting neutron star becomes unstable when accretion spins
the star just beyond this critical frequency, with an initial near balance
between viscosity and radiation reaction.  After continued spin up, however,
the radiation-reaction time can be short compared to the viscous damping time, and
the mode will then grow with a timescale of order $\tau_{GR}$
until energy loss to other modes becomes important
\cite{Bondarescu07,Bondarescu09}.
From Eqs.~(\ref{e:t_A}) and (\ref{e:tau_GR}), it follows that old neutron stars with
dominantly superconducting interiors have Alfv\'en times
shorter than the growth time of the $r$-mode. In contrast, stars with a primarily
normal interior have, by Eq.~(\ref{e:t_Anormal}),  Alfv\'en times
comparable to or longer than the radiation reaction time, if 
\begin{equation}
  B\lesssim 5\times 10^{10} f_{500}^6\,\left(\frac{M}{1.4\,M_\odot}\right)^{3/2}R^{7/2}_6 \mathrm G.
\end{equation}

\subsection{Magnetic field needed to damp the $r$-mode instability}

At first order in perturbation theory, the amplitude $\alpha(t)$ of the
unstable $r$-mode grows exponentially
\be
   \alpha(t) = \alpha(0) e^{\beta t}\,,
\ee
where $\beta = 1/\tau_{GR}$.
At second order in perturbation theory, the unstable $r$-mode has
axisymmetric differential rotation driven by a force comprising
gravitational radiation reaction and terms in
the perturbed MHD-Euler equation that are quadratic in the
first-order perturbation. The magnitude of the radiation-reaction force
per unit mass is (see, e.g., Paper I)
\be
   |{\bm f}_{GR}|\sim \alpha^2(t)\,\beta\,\Omega\, R\,.
\ee
Second-order contributions to viscous damping may reduce the
magnitude of this effective driving force; because our goal is to set an
upper limit on the second-order differential rotation, we do not include
them.  

The growth of magnetic-field energy can stop the growth of an unstable
$r$-mode when the rate at which the differential rotation increases the
energy of the {\it second-order} magnetic field,
$\langle\delta  B\rangle$,
with $\langle \cdot \rangle$ indicating the axisymmetric part of a quantity, is
equal to the rate of growth of energy of the {\it first-order} $r$-mode.

For a normal plasma, the growth rate of the magnetic-field energy
density can be roughly estimated as
\be
   \frac{d{\cal E}_m}{dt} = 4\beta{\cal E}_m \sim \frac1{2\pi} \beta\, \langle\delta B\rangle^2\,,
\label{e:dEmdt}\ee
while the energy density of the linear $r$-mode grows at the rate
\be
   \frac{d{\cal E}_{\rm mode}}{dt} = 2\beta\, {\cal E}_{\rm mode} \sim \beta\, \rho\, [\alpha(t)\, \Omega\, R ]^2\,.
\label{eq:E_mode}
\ee
The critical value of the axisymmetric part
of the perturbed magnetic field $\langle\delta B\rangle_{\rm crit}$
at which the two rates are equal is then
\be
   \langle\delta B\rangle_{\rm crit} \sim  \alpha(t)\, \Omega\, R\, \sqrt{2\pi\rho}
        \sim 10^{13}\ \alpha_{-4}\, f_{500}\,R_6\,
                \sqrt{\rho_{14.6}} \ {\rm G} \,,
\label{e:B_kritical}
\ee
where we have taken as reference saturation amplitude
$\alpha_{\rm sat} = 10^{-4}$.  As noted in Sec.~\ref{s:intro}, this
is a conservative upper limit on the
maximum value of $\alpha$ found in perturbative calculations \cite{Arras_etal03,BTW04,BTW05,Bondarescu07,BW13}), and it is much smaller
than values $\alpha_{\rm sat} \sim 10^{-1}$ - $1$ considered prior to
the perturbative papers \cite{Owen1998,Rezzolla00}.

Using the induction equation in the ideal MHD limit, it is not
  difficult to show that the secular drift of a fluid element associated
  with differential rotation in a normal core enhances an initial
  magnetic field $B_0$ by a factor of order
\be
\delta B/B_0\sim \xi^\phi,
\label{e:dB}\ee
with
$\xi^\phi$ the angular displacement of the fluid
element \cite{Rezzolla00}.  A value $\xi^\phi\gg 1$ is then needed to
amplify an initial field of $B_0 \sim 10^8 - 10^{10}$ G to the critical
value $\langle\delta B\rangle_{\rm crit} \sim 10^{13}$ G at which it can
damp or significantly alter an unstable $r$-mode.

In Sec.~\ref{s:estimates}, we will show for
an exponentially growing $r$-mode that $\xi^\phi$ has a bound of order
$  \alpha_{\rm sat}^2 \Omega/\omega_A$, which
then leads to a bound on $\delta B$. The
way it does so can be understood heuristically as follows.
Using Eq.~\eqref{e:dB}
and the expression \eqref{e:omega_A} for the Alfv\'en frequency, we can
write the perturbed magnetic field and the corresponding energy density as
\begin{eqnarray}
  \delta B &\sim& B_0\xi^\phi =  \omega_A \sqrt{\frac\rho\pi}\,R\xi^\phi,
  \nonumber\\
 \frac1{8\pi} (\delta B)^2 &\sim& \frac1{8\pi^2} \rho \omega_A^2 (R\xi^\phi)^2.
\label{e:emag}\end{eqnarray}
Then using Eq.~(\ref{e:dEmdt}) written in the form,
\be
   \frac{d{\cal E}_m}{dt} \sim \frac1{2\pi} \beta B_0^2(\xi^\phi)^2,
\label{e:dEmdt1}\ee
the bound on $\xi^\phi$ now gives
\be
   \frac{\langle\delta B_{\rm sat}\rangle}{\langle\delta B\rangle_{\rm crit}}
            \lesssim \alpha_{\rm sat},
\qquad
   \left.\frac{d{\cal E}_m/dt}{d{\cal E}_{\rm mode}/dt}\right|_{\rm sat}
        \lesssim \alpha_{\rm sat}^2,
\label{e:b_e_bound0}\ee
with numerical coefficients smaller than unity, where
$\langle\delta B_{\rm sat}\rangle$ is the magnetic field generated
by the fluid displacement $\xi^\phi$ when $r$-mode saturation occurs.

For a star with a superconducting interior, a given angular displacement
$\xi^\phi$ produces a larger magnetic energy.  However, because the
Alfv\'en frequency is correspondingly higher and $\xi^\phi$ still has a
bound of order $\alpha_{\rm sat}^2 \Omega/\omega_A$, the bound on $\xi^\phi$
is more stringent.  The net result is that the two effects cancel, and
the growth rate of magnetic energy again satisfies the bound \eqref{e:b_e_bound0}.

We can define an average perturbed magnetic field, $\langle\delta B_{\rm SC}\rangle$, as
a volume average for which $\delta {\cal E}_m = \langle\delta B_{\rm SC}\rangle^2/8\pi$.
The critical magnetic field for which the growth rate of magnetic energy and
of the linear $r$-mode are equal is then again given by Eq.~(\ref{e:B_kritical}).

\section{A toy model}
\label{s:toy}

We begin the discussion of differential rotation and magnetic field
windup with a toy model that shows the main features of the evolution of
the differential rotation and magnetic field that we claim for the
nonlinear $r$-mode. In particular, in the model, a homogeneous
  incompressible rotating fluid with cylindrical symmetry has
differential rotation driven by a force that mimics the
radiation-reaction force driving the differential rotation of the
unstable $r$-mode: It grows exponentially until a time $t_{\rm sat}$
corresponding to the saturation time of the $r$-mode and is then constant
at its final value. This limit on the growth of the driving force leads
to our main result, a stringent upper limit on the maximum angular
displacement of a fluid element and a corresponding upper limit on
magnetic field windup.  

With a driving force per unit mass having maximum magnitude ${\bm f}_{\rm max}$,
we will find an upper limit on the angular displacement $\xi^\phi_{\rm max}$
of a fluid element of order
\be
  \xi^\phi_{\rm max} \sim \frac{|{\bm f}_{\rm max}|}{R\,\omega_A^2}\,,
\ee
where $\omega_A := 2\pi/t_{A}$ is the Alfv\'en angular frequency and
$R$ the radius of the model fluid. For a normal (i.e., not
superconducting) fluid, the corresponding maximum magnetic field is of order
\be
  B_{\rm max} \sim B_0\, \xi^\phi_{\rm max}\,.
\ee

The model is essentially that introduced by Shapiro \cite{shapiro00},
differing from it only by the addition of this driving force, and, as in
Shapiro's model, the general solution to the MHD-Euler equation is
analytic. The axisymmetric, homogeneous, incompressible model
  fluid has a purely azimuthal velocity field

\be
   {\bm v}  = \Omega(t,\varpi) \bm\phi\,,
\label{e:vtoy}
\ee
where 
$\bm\phi$ is the rotational symmetry vector
\be \bm\phi =
\varpi\hat{\bm\phi} = x\hat{\bm y} - y\hat{\bm x}\,.
\ee

A magnetic field that is initially along the cylindrical radial vector
field $\hat{\bm\varpi}$, is wound up by differential rotation driven by the
exponentially growing forcing term.  With no driving force, we will see
that the dynamical equation governing the angular displacement of a fluid
element is linear, and the fluid's displacement and angular velocity can
be written as sums of normal modes with frequencies proportional to the
Alfv\'en angular frequency (\ref{e:omega_A}).

With a driving force in the azimuthal direction, differential rotation
continues to grow, and the radiation-reaction force continues to drive a
growing magnetic field. Finally, when the driving force is
time-independent (when the mode has reached saturation), the differential
rotation becomes a sum of oscillatory modes, and the magnetic field
oscillates about its
final equilibrium value.  For the nonlinear $r$-mode, the second-order
radiation-reaction force includes a part that spins down the
star. Because we are concerned here only with differential rotation, we
will restrict consideration in the toy model to a driving force that
preserves the total angular momentum of the fluid.

The toy model, like the stellar model, is governed by the MHD-Euler
system in the ideal-MHD limit, comprising the source-free Maxwell
equations and the Euler equation with a Lorentz force.  For the
incompressible fluid of the toy model, the source-free Maxwell equations
are
\be \nabla\cdot {\bm B} =0\,,
\label{e:divb_toy}
\ee
\be
   (\partial_t + \Lie_{\bm v}){\bm B}  = \partial_t {\bm B} - \nabla\times({\bm v\times B}) =0\,,
\label{e:bevol}
\ee
and the Lorentz force per unit mass is
\be
  {\bm f}_m = \frac1\rho{\bm j}\times {\bm B}
        = \frac1{4\pi\rho} (\nabla\times {\bm B})\times{\bm B}\,,
\label{eq:Lorentz}
\ee
with ${\bm j}$ the electric current density, or, equivalently
\be
f_m^i =  \frac1{4\pi\rho}  B_j(\nabla^j B^i -\nabla^i B^j)\,.
\label{eq:Lorentz_f}
\ee
With no driving force, the MHD-Euler equation has the form
\be
\bm E := \partial_t \bm v + \bm v\cdot\bm\nabla \bm v + \frac{\bm\nabla p}\rho -\bm f_m =0\,.
\label{e:MHDEulerEquation}
\ee
The differential rotation of the unstable $r$-mode is driven by the
second-order axisymmetric radiation-reaction force. This is an azimuthal
force, along $\bm\phi$, and we represent it in the toy model by a force
${\bm f}_{GR}$ per unit mass of the form
\begin{equation}
   {\bm f}_{GR} = \alpha^2(t) f(\varpi)\hat{\bm\phi}\,,
\label{e:ftoy}\end{equation}
where $f(\varpi)$ encodes the spatial dependence of the
radiation-reaction force, while the mode amplitude can be modeled simply
as given first by an exponential growth and then by a constant after time
$t_{\rm sat}$
\begin{align}
 \alpha(t) = \begin{cases}
        \alpha(0) e^{\beta t}, & t \leq t_{\rm sat} \\
                \alpha_{\rm sat} \equiv \alpha(0) e^{\beta t_{\rm sat}}\,, & t>  t_{\rm sat}.
    \end{cases}
\label{e:alphatoy}\end{align}

The time evolution of the system is then determined by
Eq.~(\ref{e:bevol}) and the driven MHD-Euler equation,
\be
 \partial_t \bm v + \bm v\cdot\bm\nabla \bm v + \frac{\bm\nabla p}\rho -\bm f_m  =\bm f_{GR},
\label{e:vevol}
\ee
with
\be
\bm\na\cdot\bm v = 0\,,
\label{e:divv}\ee
because of the incompressibility assumption. Equation~(\ref{e:divv}) is
identically satisfied by a velocity field of the form (\ref{e:vtoy}), and
the evolution equation for ${\bm B}$, Eq.~(\ref{e:bevol}), keeps ${\bm B}$
divergence free.

The $\varpi$ and $z$ components of Eq.~(\ref{e:bevol}) are
$\partial_t B^\varpi = \partial_t B^z = 0$. The model has vanishing $B^z$,
and Eq.~(\ref{e:divb_toy}) implies that $B^\varpi$ has the temporally constant form
\be
   B^\varpi = \frac R\varpi B_0\,,
\ee
where $R$ is the radius of the cylinder.

Only the $\phi$ component of the magnetic field is dynamical, and it is
expressed in terms of $\xi^\phi$ by a first integral of the $\phi$
component of Eq.~(\ref{e:bevol}), namely
\be
   B^\phi = \frac{R}{\varpi} B_0\, \pa_\varpi\, \xi^\phi\,.
\label{e:bphi}
\ee
Hence, for a stationary system
  in which $\Omega(\varpi)=\partial_t\xi^\phi$ is constant in time,
$B^{\phi}$ will simply grow linearly in
  time; this is the well-known magnetic-field ``winding,'' producing a
  toroidal magnetic field out of a purely poloidal one
  \cite{ah63,Spruit99,Rezzolla00}.

As the instability develops and saturates, however, the evolution of
the angular displacement $\xi^\phi$ is given by the $\phi$ component
of Eq.~(\ref{e:vevol})
\be
  \partial_t^2 \xi^\phi
     - \omega_A^2 \frac{R^4}{4\pi^2\varpi^3}\partial_\varpi(\varpi\partial_\varpi \xi^\phi)
   = \alpha^2(0) f e^{2\beta t}\,,
\label{e:eulerphi}
\ee
where $\omega_A$ is given by Eq.~(\ref{e:omega_A}).
Two remarks are worth making about
  Eq. (\ref{e:eulerphi}). First, it has a simple mechanical equivalent in
  terms of a driven harmonic oscillator, whose driving force first grows
  exponentially and then becomes time independent after $t_{\rm sat}$. 
Second, although it is derived from the MHD-Euler equation, it
  does not involve the pressure: The remaining $\varpi$
component of the MHD-Euler equation determines $p$ but is not needed
for the evolution of $\xi^\phi$, ${\bm B}$ or $\Omega$.

We model crust pinning of the magnetic field by the boundary condition
\be
 B^\phi(\varpi=R) =0\,,
\ee
and Eq.~(\ref{e:bphi}) then implies
\be
  \partial_\varpi \xi^\phi(\varpi=R) = 0\,.
\ee

Setting ${\mathfrak r}:=\varpi^2/R^2$ allows us to write the homogeneous MHD-Euler equation in
the form of a cylindrical wave equation $\partial_t^2\xi^\phi -
\pi^{-2}\omega_A^2 {\mathfrak r}^{-1}\partial_{\mathfrak r} ({\mathfrak r}\,
\partial_{\mathfrak r} \xi^\phi) =0$, whose solutions are proportional to
Bessel functions of order 0,
\be
   \xi_n^\phi = J_0(k_n\varpi^2/R^2) e^{i\omega_n t}\,,
\ee
where $k_n$ is the $n$th zero of $J_0'$ and
\be
    \omega_n =  \omega_A k_n/\pi\,.
\ee
Writing $f$ and $\xi^\phi$ as sums of orthogonal eigenfunctions
\begin{subequations}
\begin{align}
f &= \sum f_n J_0(k_n \varpi^2/R^2)\,, \\
\xi^\phi &= \sum c_n(t) J_0(k_n \varpi^2/R^2)\,,
\end{align}
\end{subequations}
we obtain the exponentially growing solution to Eq.~\eqref{e:eulerphi}
prior to $t_{\rm sat}$,
\be
\xi^\phi
   = \sum_n\frac{\alpha^2(t)}{4\beta^2+\omega_n^2}f_n J_0(k_n\varpi^2/R^2).
\label{e:enorm}
\ee

On the other hand, when the driving force is time independent,
representing an $r$-mode after nonlinear saturation is reached,
the term $\alpha^2(0) f e^{2\beta t}$ in Eq.~(\ref{e:eulerphi}) is replaced
by the time-independent term $\alpha_{\rm sat}^2 f$.  The solution for $\xi^\phi$
is now the sum of a time-independent term
and harmonic functions of angular frequency $\omega_n$
With a constant force, the equilibrium value of $\xi^\phi$ is
obtained by omitting $4\beta^2$ from the denominator of Eq.~(\ref{e:enorm}), and $\xi^\phi$
has the form
\begin{eqnarray}
  \xi^\phi &=& \sum_n\frac{\alpha_{\rm sat}^2 }{\omega_n^2}f_n J_0(k_n\varpi^2/R^2)
  \nonumber\\
&&        + \sum_n a_n J_0(k_n\varpi^2/R^2) \cos(\omega_n t+\eta_n)\,.
\label{e:xi_sat}
\end{eqnarray}

The amplitude of the harmonic term depends on the transition from exponential growth to
a time-independent driving force.  A gradual approach to saturation reduces the amplitude, and we
set an upper limit by adopting a driving force whose growth stops instantaneously,
as given by Eqs.~\eqref{e:ftoy} and \eqref{e:alphatoy}.

The oscillation amplitude is then
\be
  a_n = \alpha_{\rm sat}^2\frac{2\beta}{\omega_n^2\sqrt{4\beta^2+\omega_n^2}}f_n
    < \frac{\alpha_{\rm sat}^2}{\omega_n^2} f_n,
\ee
implying a maximum value of $\xi^\phi$ less than twice its equilibrium value.
Here we have assumed that, prior to $t_{\rm sat}$, $\xi^\phi$ is dominated by
the exponentially growing solution (\ref{e:enorm}) associated with the unstable $r$-mode.

Equations~(\ref{e:enorm}) and (\ref{e:xi_sat}) give us the
toy-model's exact expressions for the angular displacement of a fluid
element. We now consider its implications for the unstable $r$-mode,
assuming that the behavior of the toy model's differential rotation is
similar to that of the $r$-mode. The axisymmetric part of the $r$-mode's
radiation-reaction force per unit mass is of order [cf.,
  Eq. \eqref{eq:E_mode}]
\be
   \langle |{\bm f}_{GR}|\rangle \sim \alpha^2(t)\, \beta\,\Omega\, R\,.
\ee

For a normal interior, the Alfv\'en frequency (\ref{e:omega_A}) has
magnitude
\begin{equation}
  \omega_{A} = 0.9\times 10^{-4}\ B_9\, R_6^{-1}\,\rho_{14.6}^{-1/2}\,.
\label{eq:omega_A_normal}
\end{equation}
With $f_n \sim \beta\,\Omega$ and hence $\alpha^2(t)\, R\, f_n$ of
order $|{\bm f}_{GR}|$ and decreasing for large $n$, the sum in
Eq.~(\ref{e:enorm}) is dominated by modes with $k_n\sim 1$ and
$\omega_n\sim\omega_A$.  Prior to saturation, we then have a bound that is independent of
$\beta$,
\be
 \xi^\phi_{\rm sat} \lesssim \alpha_{\rm sat}^2 \frac{\beta\Omega}{4\beta^2+\omega_A^2}
       < \alpha_{\rm sat}^2 \frac{\Omega}{4\omega_A},
\label{e:xi_bound}\ee
implied by the relation
\be
  \frac{\beta}{4\beta^2+\omega_A^2} = \frac1{4\omega_A}\left[1- \frac{(2\beta-\omega_A)^2}{4\beta^2+\omega_A^2}\right]
    \leq \frac1{4\omega_A}.
\label{e:ineq}\ee
From Eq.~(\ref{e:bphi}), an angular displacement $\xi^\phi$ with characteristic wavelength
of order $R$ gives a magnetic field  $B^{\hat\phi} \sim B_0 \xi^\phi$, with a
corresponding upper limit prior to saturation
\begin{align}
   B^{\hat\phi}_{\rm sat} & \sim \xi^\phi B_0 \nonumber \\
    & \lesssim \alpha_{\rm sat}^2 \frac\Omega{4\omega_A} B_0
       = \frac1{4\sqrt\pi}\alpha_{\rm sat}^2 \Omega R \rho^{1/2} .
\nonumber\end{align}
or
\be
  B^{\hat\phi}_{\rm sat} \lesssim \frac1{4\sqrt\pi} \alpha_{\rm sat} B_{\rm crit},
\label{e:Bsat_limit}\ee
where we have used Eq.~(\ref{e:omega_A}) for $\omega_A$ and Eq.~(\ref{e:B_kritical})
for the critical magnetic field needed to damp the $r$-mode.
The corresponding inequality for the change in the magnetic energy density
at quadratic order in $\xi^\phi$ is
\be
  \delta {\cal E}_{\rm sat} \lesssim \frac1{16\pi} \alpha_{\rm sat}^2 \delta {\cal E}_{\rm crit}
\ee
Then $\alpha_{\rm sat}\ll 1$ implies $B^{\hat\phi}_{\rm sat} \ll B_{\rm crit}$,
or ${\cal E}_{\rm sat} \ll {\cal E}_{\rm crit}$.
This is our main result.

After saturation, the linear $r$-mode is no longer growing.  Energy gained from the
first-order radiation reaction is balanced by energy loss to daughter modes
and to dissipation, and we now ask whether magnetic-field windup can
play a significant role at this stage.   In the post-saturation evolution
of the angular displacement given by Eq.~(\ref{e:xi_sat}), $\xi^\phi$ reaches
and oscillates about an equilibrium value that can be large if
$\omega_A$ is small.  That is, from Eq.~(\ref{e:xi_sat}), we have
\be
   \xi^\phi \lesssim \alpha_{\rm sat}^2 \frac{\beta\Omega}{\omega_A^2},
\label{e:xi^phi}\ee
\be
  B^{\hat\phi} \lesssim \alpha_{\rm sat}^2 \frac{\beta\Omega}{\omega_A^2} B_0.
\label{e:Bpostsat}\ee
Now, however, the growth rate of each mode is proportional to $\omega_n$.
Eq.~(\ref{e:dEmdt}) is then replaced by
\be
   \frac{d{\cal E}_m}{dt} \sim \omega_A\, \langle\delta B\rangle^2\,,
\ee
and the critical magnetic field for which the energy gained from radiation reaction is
comparable to the energy lost to magnetic-field windup is given by
\bea
\langle\delta B\rangle_{\rm crit} &\sim& \alpha_{\rm sat} \Omega R\sqrt{\rho\beta/\omega_A}
\nonumber\\
   &\geq& 1.5\times10^{13}\alpha_{-4}\beta_{-3.3}^{1/2}f_{500} R_6^{3/2}\rho_{14.6}^{3/4}B_9^{-1/2}\ G,\nonumber\\
\label{e:Bcritpost}\eea
where we have used $\omega_A\geq \omega_A(B_0) = (B_0/R)\sqrt{\pi/\rho}$.  Eqs.~(\ref{e:Bpostsat}) and
(\ref{e:Bcritpost}) imply
\bea
  \frac{B^{\hat\phi}}{\langle\delta B\rangle_{\rm crit}}
    &\lesssim& \frac{\alpha_{\rm sat}}{\pi^{3/4}}
  \beta^{1/2}R^{1/2}\rho^{1/4} B_0^{-1/2}
  \nonumber\\
        &=& 1.3 \times 10^{-4} \alpha_{-4} \beta_{-3.3}^{1/2}R_6^{1/2}
                \rho_{14.6}^{1/4} B_9^{-1/2}\ .\quad
\label{e:Bpost_limit}\eea
To reach the critical magnetic field, one would need a normal interior with
$B_0$ of order $20$ G, more than 6 orders of magnitude smaller than the smallest
estimated external magnetic field in an old neutron star
($4.5\times 10^7$ G, inferred from the period and spin-down of PSR J1938+2012 \cite{Stovall_etal16}).%
\footnote{Although interior fields below 100 G seem highly unlikely,
field decay to that level has not, to our knowledge, been ruled out
observationally.}
Equation~\eqref{e:Bpost_limit} implies that the post-saturation growth
of an initial magnetic field of $10^8$ or $10^9$ G will continue to
satisfy the saturation constraint \eqref{e:Bsat_limit}.

The growth of a realistic initial magnetic field is then much too
small to alter the $r$-mode. In particular, for neutron star whose
interior is a normal plasma, the maximum angular displacement is of order
\be
  \xi^\phi_{\rm max}
        \sim 2\ \alpha_{-4}^2\beta_{-3.3}f_{500} \rho_{14.6}R_6^2B_9^{-2}\ \rm rad,
\label{e:xi_toy_normal}\ee
and a corresponding maximum change in the magnetic field is
\be
   B^{\hat\phi}_{\rm max} \sim \xi^\phi B_0
    \lesssim 2\times 10^9 \alpha_{-4}^2\beta_{-3.3}f_{500} \rho_{14.6}R_6^2B_9^{-1} \rm \ G,
\label{e:B_toy_limit}\ee
as implied by Eq.~\eqref{e:Bpostsat}.

Again, two remarks are in order here.
First, because Eq. \eqref{e:xi^phi} refers to a time after
saturation has been achieved, the azimuthal displacement in
\eqref{e:xi_toy_normal} has a time-independent equilibrium value.
Using again the mechanical equivalent discussed above, such a time-independent
displacement corresponds to that of a harmonic
oscillator subject to a constant and time-independent gravitational
force.
Second, in this toy model, because the poloidal component $B^\varpi$
is constant and decoupled from the growth of the toroidal field,
the frequencies $\omega_n$ of the modes are constant in time:
They do not grow with the growth of the toroidal field.
As a result, the quadratic dependence on the mode's amplitude in
Eq. \eqref{e:B_toy_limit} can increase the magnetic field by six or
more orders of magnitude if $\alpha_{\rm sat} \simeq 0.1-1$, as was
assumed in earlier work \cite{Owen1998,Rezzolla00}.

The exact decoupling that keeps $\omega_n$ constant may be an artifact of
the toy model: Sec.~\ref{s:2ndorder} displays the second-order MHD-Euler
equations governing differential rotation generated by an unstable $r$-mode.
In this more realistic model, we have checked that, for a generic background
magnetic field, there is no analogous decoupling of poloidal and toroidal fields.
Nevertheless, numerical evolutions of the MHD-Euler equations \cite{schr13}
show a poloidal field whose
magnitude remains approximately constant while differential rotation
winds up the magnetic field. We therefore do not assume that an increasing
magnetic field produced by differential rotation results in an increased
frequency of modes associated with the field windup.

For cold neutron stars whose interior is a type II superconductor,
  we find in Sec. V that an essentially equivalent version of the
  constraint~\eqref{e:Bsat_limit} holds both before {\it and} after
  nonlinear saturation.  Before encountering the detailed calculation in
  Sec. V, we can understand the result heuristically as follows.  The
  energy density of a stellar mode with displacement $\xi$ and frequency
  $\omega$ is of order $\rho \omega^2\xi^2$.  In particular, differential
  rotation drives modes whose energy density is of order ${\delta\cal E}
  \sim \rho \omega_A^2 \xi^2$, where $\xi\sim R\xi^\phi$.  The Alfv\'en
  frequency $\omega_{A,SC}$ of a superconducting interior is much larger
  than that of a normal plasma, and the rate of growth of magnetic energy
  is thus much larger for a given displacement $\xi^\phi$. However,
  because the bound $\xi^\phi \lesssim \alpha_{\rm
    sat}^2\Omega/\omega_{A,SC}$ on $\xi^\phi$ is more stringent by the
  factor $\omega_A/\omega_{A,SC}$, the bound on ${\delta\cal E}_m$
  remains the same: \bea {\delta\cal E}_m &\lesssim& \alpha_{\rm
    sat}^4\rho\Omega^2 R^2 \nonumber\\
  &\sim& \alpha_{\rm sat}^2(\textrm{energy density
    of the linear $r$-mode}).\qquad  \eea The constraint also holds after
  saturation because, as we noted in Sec.~\ref{s:timescales},
  $\omega_{A,SC}\gg \beta$, implying that the equilibrium displacement is
  within about a factor of two of the displacement at saturation.  We
  conclude that, for small saturation amplitudes ($\alpha_{\rm
    sat}\lesssim 10^{-4}$), magnetic field windup from differential
  rotation is too small to produce magnetic fields that can damp or
  significantly alter the unstable $r$-mode.

\section{Equilibrium and perturbation equations}
\label{s:perturbation}

We work in the approximation of Newtonian MHD
with the star described
by a perfect fluid with infinite conductivity.
The version of the Euler equation that we use,
Eq.~(\ref{e:FullEulerEquation}), includes ${\bm f}_{GR}$, the
post-Newtonian gravitational radiation-reaction force (per unit mass).
This force plays a central role in the nonlinear evolution of the
$r$-modes that is the primary focus of our paper. Because the old neutron
stars we consider have spin-down times much longer than the
gravitational radiation-reaction timescale of an $r$-mode (and may also
be balanced by accretion), we neglect radiation reaction associated
with the magnetic field.

We denote by $Q:=\{\rho, {\bm v}, p,\Phi, {\bm B}, {\bf E}\}$ the
collection of fields that determine the state of the fluid. Here $\rho$
is the mass density, $v^i$ the fluid velocity, $p$ the pressure, $\Phi$
the gravitational potential, and ${\bf E}$ and ${\bm B}$ the electric
and magnetic fields.  For a barotropic equation of state, $p=p(\rho)$,
the specific enthalpy $h$ of the fluid is
\be
h= \int_0^p \frac{dp}{\rho}\,,
\ee
and we define a potential $U$ by
\be
U:= h + \Phi\,,
\ee
where $\Phi$ satisfies the Poisson equation
\be
\nabla^2\Phi = 4\pi\rho\,.
\label{e:GravPotentialEq}
\ee

The following equations govern the evolution of the fluid and its
electromagnetic field.  With $g_{ij} $ the flat 3-metric and $g$ its
determinant, conservation of mass (the continuity equation) has the
equivalent forms
\be
  (\partial_t + \Lie_{\bm v} )(\rho\sqrt g) = 0
                        = \partial_t \rho + \bm\nabla\cdot(\rho\bm v)\,,
\label{e:masscon}
\ee
where we use the relation $\frac1{\sqrt g} \Lie_{\bm v}\sqrt g = \nabla\cdot {\bm v} $.
The vanishing of the electric field in a comoving frame is given by
\be
  \bm E + {\bm v} \times {\bm B} = 0\,,
\ee
\be
(\partial_t + \Lie_{\bm v})({\bm B}\sqrt g) = 0 = \partial_t {\bm B} - \nabla\times({\bm v} \times{\bm B})\,,
\label{e:dbdt}
\ee
and by expression \eqref{eq:Lorentz} for the Lorentz force per unit
  mass.  Recalling that ${\bm f}_{GR}$ is the radiation-reaction force
per unit mass associated with gravitational radiation, we write the MHD-Euler
equation in the form
\be
  \bm  \partial_t \bm v + \bm v\cdot\bm\nabla \bm v + \bm\nabla U -\bm f_m = \bm f_{GR}\,.
\label{e:FullEulerEquation}
\ee

The radiation-reaction force per unit mass is given by \cite{fll15,Lindblom2001,Rezzolla1999}
\bea
&&
\!\!\!\!\!\!\!\!\!
{\bm f}_{GR}=\sum_{l\geq 2}\sum_{|m|\leq l} \frac{(-1)^{\ell+1}N_\ell}{32\pi}
\, \Re\Biggl\{\,
\frac{{\bm\nabla}(r^\ell Y^{\ell m})}{\sqrt{\ell}}
\frac{d^{\,2\ell+1}I^{\ell m}}{dt^{\,2\ell+1}}\nonumber\\
&&
\!\!\!\!\!
-\frac{2r^\ell\bm Y^{\ell m}_{B}}{\sqrt{\ell+1}}
\frac{d^{\,2\ell+2}S^{\ell m}}{dt^{\,2\ell+2}}
-
\frac{2{\bm v} \times \bm \nabla(r^\ell\, Y^{\ell m})}{\sqrt{\ell}}
\frac{d^{\,2\ell+1}S^{\ell m}}{dt^{\,2\ell+1}}\Biggr\},\nonumber\\
\label{e:GRForceDef}
\eea
where $\Re (Z)$ denotes the real part of the complex quantity $Z$. The
quantities $I^{\ell m}$ and $S^{\ell m}$ are the complex mass and current
multiple moments of the fluid source (cf. Thorne~\cite{Thorne1980}
Eqs.~5.18a,b) defined by,
\bea
I^{\ell m}&:=&
\frac{N_\ell}{\sqrt{\ell}}
\int \rho\, r^\ell Y^{*\ell m} d^3 x,\\
S^{\ell m}&:=&\frac{2N_\ell}{\sqrt{\ell+1}}
\int \rho \,r^\ell\, {\bm v}  \cdot \bm Y^{*\ell m}_{B} d^3x,
\eea
with $N_\ell$ the constant
\bea
N_\ell := \frac{16\pi}{(2\ell+1)!!}
\sqrt{\frac{(\ell+2)(\ell+1)}{2(\ell-1)}}.
\label{e:NlDef}
\eea
The functions $Y^{\ell m}$ are the standard spherical harmonics,
while the $\bm Y^{\ell m}_B$ are the magnetic-type vector
harmonics defined by
\bea
\bm Y^{\ell m}_B := \frac{\bm r\times \bm\na Y^{lm}}{\sqrt{\ell(\ell+1)}}\,,
\eea
with normalization $\int |Y^{\ell m}|^2 d\cos\theta\, d\phi = 1$ and $
\int |\bm Y^{\ell m}_B|^2 d\cos\theta\, d\phi = 1$. In Cartesian
coordinates, $\bm r$ is given by $\bm r =(x,y,z)$.

\subsection{Equilibrium equations}

We consider a uniformly rotating, axisymmetric equilibrium star with angular velocity $\Omega$.
Because the magnetic field is not in general aligned with the axis of symmetry,
the equilibrium is stationary only in a rotating frame, satisfying
\be
  (\partial_t + \Lie_{\bm v}) Q = 0\,,
\ee
where
\be
   {\bm v}  = \Omega\bm\phi\,,
\ee
with $\bm\phi$ the generator of rotations about the $z$-axis. In Cartesian coordinates,
$\bm\phi = (-y,x,0)$, implying
\be
   \bm\phi\cdot\bm\phi = \varpi^2\,,
\ee
where $\varpi$ is the distance from the rotation axis.

We consider constant-mass sequences of stellar models, i.e., models
whose exact mass perturbations, $\delta M = M(\alpha)- M(\alpha=0)$
vanish identically for all values of
$\alpha$. The integrals of the $n^\mathrm{th}$-order density
perturbations therefore vanish identically for these models:
\be
0=\frac{1}{n!}\left.\frac{d^{\,n} M(\alpha)}{d\alpha^n}\right|_{\alpha=0} =
\int \deltaN \rho\, \sqrt{g}\,d^{\,3}x.
\label{e:MassConservationIntegral}
\ee

From Eq.~(\ref{e:FullEulerEquation}) with ${\bm f}_{GR}=0$, the Euler equation
governing the equilibrium is
\be
  \nabla_i (U-\frac12\varpi^2\Omega^2) + \frac1{4\pi\rho} B^j(\nabla_i B_j - \nabla_j B_i) =0\,,
\ee
where we have used the relation $(\partial_t + \Lie_{\bm v})v_i = 0$.

\subsection{Eulerian and Lagrangian perturbations}
\label{s:perturbations}

We denote by $Q(\alpha,t,x)$ a one-parameter family of stellar
models. For each value of the parameter $\alpha$, $Q(\alpha,t,x)$
satisfies the full nonlinear time-dependent
Eqs.~(\ref{e:GravPotentialEq})--(\ref{e:FullEulerEquation}). The amplitude $\alpha$
is time independent and can be identified with the initial amplitude $\alpha(0)$
when we describe a growing mode by a time-dependent $\alpha(t)$.

The exact Eulerian perturbation $\delta Q$, defined as the difference between
$Q(\alpha)$ and $Q(0)$, is defined everywhere on the intersection of
the domains where $Q(\alpha)$ and $Q(0)$ are defined as
\bsube
\bea
\delta Q(\alpha, t,x)&:=& Q(\alpha,t,x)-Q(0,t,x)\\
&=& \alpha\, \deltaI Q(t,x) + \alpha^2\, \deltaII Q(t,x) + {\cal O}(\alpha^3)\,,
\nonumber\\\label{e:delta Q}
\eea
\esube
where the n-th order perturbation $\deltaN Q$ is
\be
\deltaN Q(t,x) := \frac{1}{n!}\frac{\partial^{\,n}\,
Q(\alpha,t,x)}
{\partial\alpha^n}\biggr|_{\alpha=0}\,.
\label{e:deltaN Q Def}
\ee
Although the exact Eulerian perturbation has meaning only on the intersection of
the support of the unperturbed and perturbed fluid, $\deltaN Q$ is well defined
everywhere in the interior of the unperturbed star.

Exact Lagrangian perturbations can be defined by introducing a
diffeomorphism $\chial$ that maps fluid elements in the equilibrium star
$Q(0,t,x)$ to the corresponding elements in the solution
$Q(\alpha,t,x)$. The exact Lagrangian change in a quantity $Q$ is defined
by,
\bea
\Delta Q(\alpha,t,x) &:=& \chial^* Q(\alpha,t,x)
- Q(0,t,x)
\label{e:LagrangianDelta1}\\
&=& \alpha \DeltaI Q + \alpha^2 \DeltaII Q + {\cal O}(\alpha^3)\,,
\label{e:LagrangianDelta2}
\eea
where $\chi^*_\alpha$ is the pullback map (see Appendix \ref{s:LagrangianPerturbations})
and
\be
\DeltaN Q(t,x)
:= \frac{1}{n!}\frac{\partial^{\,n}\, \chial^* Q(\alpha,t,x)}
{\partial\alpha^n}\biggr|_{\alpha=0}\,.
\ee
We can write $\Delta Q$ in terms of a Lagrangian perturbation vector $\xi^i$
in the manner
\bea
\Delta Q(\alpha,t,x)
&=& \left(1 + \Lie_{\xi}+\half\Lie_{\xi}^2\right)
\Big[Q(0,t,x)+\delta Q(\alpha,t,x)\Big]\nonumber\\
&&-Q(0,t,x) + \mathcal{O}(\alpha^3)\,.
\eea
With
\be
\xi^i = \alpha \xiI^i + \alpha^2 \xiII^i + {\cal O}(\alpha^3)\,,
\label{e:LagrangianExpansion}
\ee
the first- and second-order Lagrangian perturbations are given by
[see Appendix \ref{a:2nd_order_displacement}, Eq. (\ref{e:DeltaIDeltaII})],
\bsube
\bea
&&\!\!\!\!\!\!\!
\DeltaI Q(t,x)=\left(\deltaI+\Lie_{\xiI}\right)Q(0,t,x),  \label{e:DeltaIQ}\\
&&
\!\!\!\!\!\!\!
\DeltaII Q(t,x)=\nonumber\\
&&\quad
\left[\deltaII+\Lie_{\xiII}
+\Lie_{\xiI}\deltaI+\half\Lie_{\xiI}^2
\right]Q(0,t,x)\,.\qquad
\label{e:DeltaIIQ}\eea\label{e:DeltaQ}
\esube
The components of the vectors $\xiI^i$ and $\xiII^i$ are given in
any coordinates by
\bea
\xiI^i &=& \frac{\partial \chial^i}{\partial \alpha}\biggr|_{\alpha=0},
\label{e:Lagrangian1}\\
\xiII^i &=&  \frac{1}{2}
\left.\frac{\partial^2 \chial^i}{\partial\alpha^2}\right|_{\alpha=0}
\!\!-\,\,\frac12\xiI^j\partial_j \xiI^i\,.
\label{e:Lagrangian2}
\eea

 The commutator 
\[
   \Delta (\partial_t + \Lie_{\bm v} ) = (\partial_t + \Lie_{{\bm v} _0}) \Delta\,
\]
obtained as Eq.~\eqref{e:DeltaLcommute} of Appendix~\ref{s:LagrangianPerturbations}, 
gives the perturbed mass-conservation equation (\ref{e:masscon}) and induction equation (\ref{e:dbdt})
in the forms
\bsube
\bea (\partial_t+\Lie_{\bm v_0})\Delta(\rho\sqrt g) &=& 0\,, \\
    (\partial_t+\Lie_{\bm v_0})\Delta(B^i\sqrt g) &=& 0\,,
\eea
\esube
where $v_0$ is the unperturbed velocity field and $\Delta$ is the exact Lagrangian perturbation.  These equations have first integrals
\bsube
\bea
  \Delta(\rho\sqrt g) = 0\,, \\
  \Delta(B^i\sqrt g) = 0\,,
\eea
\esube
correct to all orders in $\alpha$, implying
\be
  \Delta\frac{\bm B}\rho = 0\,.
\label{e:DeltaBrho}\ee

The first- and second-order Lagrangian perturbations of $g_{ij} $ and $\sqrt g$ are given by
\bsube\bea
  \DeltaI g_{ij}  &=& 2\nabla_{(i} \xiI_{j)}\,,\\
  \DeltaII g_{ij}  &=& 2\nabla_{(i} \xiII_{j)} +\xiI^k\na_k\na_{(i}\xiI_{j)}
  \nonumber\\
   &&\,\,+\na_i\xiI^k\na_j\xiI_k +\na_k\xiI_{(i}\na_{j)}\xiI^k\,,\qquad
\eea\esube
\bsube
\bea
&&\!\!\!\!\!\!\!\!\!\!\!\!
\frac1{\sqrt g} \DeltaI \sqrt g = \bm\na\cdot{\bm\xi}^{(1)}\,,\\
&&\!\!\!\!\!\!\!\!\!\!\!\!
\frac1{\sqrt g}\DeltaII \sqrt g = 
  \nonumber\\
  &&\,\,\bm\na\cdot{\bm\xi}^{(2)}
  +\frac12 (\bm\na\cdot{\bm\xi}^{(1)})^2
        +\frac12\bm\xiI\cdot\bm\na\bm\na\cdot{\bm\xi}^{(1)}\,,\quad
\eea\esube
and the corresponding perturbations of $\rho$ and $p$ are
\bsube\begin{align}
  \frac{ \DeltaI \rho}{\rho} =& -\bm\na\cdot{\bm\xi}^{(1)},\\
  \frac{\DeltaII \rho}{\rho} =& -\bm\na\cdot{\bm\xi}^{(2)} \nonumber\\
                  &+\frac12(\bm\na\cdot{\bm\xi}^{(1)})^2
                  -\frac12\bm\xiI\cdot\bm\na\bm\na\cdot{\bm\xi}^{(1)}\,,                  
\end{align}\label{e:Delta_rho}\esube
\bsube
\bea
  \frac{ \DeltaI p}{\gamma p}  &=& -\bm\na\cdot{\bm\xi}^{(1)}\,,\\
  \frac{\DeltaII p}{\gamma p} &=& -\bm\na\cdot{\bm\xi}^{(2)}  +\frac12\left(\gamma
    + \frac{\partial\log\gamma}{\partial\log\rho}\right) (\bm\na\cdot{\bm\xi}^{(1)})^2
\nonumber\\
    &&    -\frac12\bm\xiI\cdot\bm\na\bm\na\cdot{\bm\xi}^{(1)}\,,
\eea\label{e:Delta_p}\esube
with $\gamma=d\log p/d\log \rho$ the adiabatic index.

The first- and second-order Lagrangian perturbations of the covariant and contravariant
forms of the magnetic field are then
\bsube
 \begin{align}
 \DeltaI B^i =& - B^i\nabla_j\xiI^j\,, \\
 \DeltaII B^i =& -B^i\nabla_j\xiII^j\nonumber\\
 & \!\!\!\!
 +B^i\left[\frac12(\nabla_j\xiI^j)^2 - \frac12\xiI^k\nabla_k\nabla_j\xiI^j\right],
\label{e:DeltaIB}\end{align}\esube
and
\bsube
 \begin{align}
 \DeltaI B_i &= B^j[2\na_{(i}\xiI_{j)} - g_{ij} \nabla_k\xiI^k]\,,
\label{e:DeltaIB_i}\\
\DeltaII B_i &= B^j[2\na_{(i}\xiII_{j)}-g_{ij} \na_k\xiII^k]
\nonumber\\
&    +B^j\left[\xiI^k\na_k\left(\na_{(i}\xiI_{j)}-\frac12g_{ij} \na_l\xiI^l\right)\right.
\nonumber\\
&\ \ +\na_i\xiI^k\na_j\xiI_k +\na_k\xiI_{(i}\na_{j)}\xiI^k
\nonumber\\
&\left. \ -2\na_{(i}\xiI_{j)}\na_k\xiI^k+\frac12 g_{ij} \left(\na_k\xiI^k\right)^2
\right].
\label{e:Delta_B_i}\end{align}\label{e:Delta_B}\esube

Finally, the expressions for the Lagrangian changes in the contravariant and covariant velocity
are (see Appendix \ref{s:Delta_v})
\bsube 
\bea
\DeltaI v^i &=& \partial_t \xiI^i\,,\\
\DeltaII v^i &=& \partial_t \xiII^i + \half\Lie_{\xiI}\partial_t\xiI^i\,,
\label{e:Delta_v^i}\eea\esube
implying
\bsube\bea
\DeltaI v_i &=& \partial_t \xiI_i + 2\nabla_{(i}\xiI_{j)} v^j\,,\\
\DeltaII v_i &=& \partial_t \xiII_i + 2\nabla_{(i}\xiII_{j)} v^j
+\pa_t\xiI^j\na_i\xiI_j \nonumber\\  
&&+ \half\partial_t(\xi^{(1)j}\na_j\xiI_i)
+(\xi^{(1)k}\na_k \nabla_{(i}\xiI_{j)}\nonumber\\
&&+ \na_k \xiI_{(i}\na_{j)}\xi^{(1)k} 
        +\na_i\xiI_k\na_j\xi^{(1)k}v^j\,.\qquad  
\label{e:Delta_v_i}\eea\esube

\subsection{First-order perturbation equations}

We now consider perturbations of the MHD-Euler system, at first order in the amplitude
$\alpha$. We use the formalism of Friedman and Schutz \cite{fs78} and its
extension to the MHD-Euler system by Glampedakis and Andersson \cite{GA07}.
To write the perturbed MHD-Euler equation (\ref{e:FullEulerEquation}),
\begin{align}\!\!\!\!\!
 \rho\DeltaI \mathpzc E_i :=& \rho\DeltaI\biggl[(\partial_t + v^j\nabla_j) v_i + \frac{\nabla_i p}\rho + \nabla_i \Phi\nonumber\\
&\ \ + \frac1{4\pi\rho}B^j(\nabla_i B_j - \nabla_j B_i)\biggr]
= \rho\deltaI f_{GR\,i}\,,\!
\label{e:DeltaIE_i}\end{align}
in terms of the Lagrangian displacement $\bm\xi^{(1)}$, we use the first-order part of Eq.~(\ref{e:DeltaBrho}),
\be
   \DeltaI\frac {B^i}\rho = 0\,,
\ee
and obtain for the term involving the perturbed Lorentz force the form
\bea
&&   \rho\DeltaI\left[\frac1{4\pi\rho} B^j(\nabla_i B_j - \nabla_j B_i)\right]
\nonumber\\
    &&\qquad= \frac1{4\pi} B^j(\nabla_i \DeltaI B_j - \nabla_j \DeltaI B_i)
\nonumber\\
    &&\qquad=\frac1{2\pi}
        B^j\Bigl[
          \na_i(B^k\na_{(k}\xiI_{j)}) - \na_j(B^k\na_{(k}\xiI_{i)})
          \qquad\nonumber\\
&&\qquad\qquad\qquad                   -\na_{[i}(B_{j]}\nabla_k\xiI^k)
            \Bigr]\,,\qquad
\eea
where we have used Eq.~(\ref{e:DeltaIB_i}) and the fact that Lie and exterior derivatives commute.

The perturbed MHD-Euler equation (\ref{e:DeltaIE_i}) has the form
\be
  A_{ij} \partial_t^2 \xiI^j + B_{ij} \partial_t\xiI^j +  C_{ij}  \xiI^j
        = \rho\deltaI f_{GR\,i}\,,
\label{e:DeltaIE_i2}\ee
where
\bsube
\bea
  A_{ij}  &:=&\rho g_{ij} \,,
\label{e:Aab}\\
  B_{ij}  &:=& 2\rho g_{ij} v^k\nabla_k \,,
\label{e:Bab}\\
C_{ij} \xi^j &:=& \rho (v^j\nabla_j)^2\xi_i - \nabla_i(\gamma p \nabla_j\xi^j)+ \nabla_i p \nabla_j\xi^j 
\nonumber\\
&&-\nabla_j p \nabla_i \xi^j+\rho\xi^j\nabla_j\nabla_i\Phi + \rho \nabla_i\deltaI\Phi
\nonumber\\
   &&+ \frac1{2\pi} B^j\Bigl[
            \na_i(B^k\na_{(k}\xi_{j)}) - \na_j(B^k\na_{(k}\xi_{i)})\nonumber\\
&&\quad                  -\na_{[i}(B_{j]}\nabla_k\xiI^k)
            -\na_i\xiI^k  \nabla_{[k}B_{j]}
            \Bigr] \,.\qquad
\label{e:Cab}\eea\label{e:ABC}
\esube
Here $\deltaI\Phi$ is the asymptotically vanishing solution to the perturbed Poisson equation
\[
   \nabla^2 \deltaI\Phi = 4\pi\deltaI\rho = -4\pi \bm\na\cdot(\rho \bm\xi^{(1)})\,.
\]
For vectors $\xi^i$ and $\eta^i$ that vanish at the boundary of the star, the operators
$A_{ij} $ and $C_{ij} $ are self-adjoint in the sense
\bea
  \int dV \eta^i C_{ij}  \xi^j = \int dV \xi^i C_{ij}  \eta^j\,,
\eea
and $B_{ij} $ is anti-self-adjoint.

The exact perturbed gravitational radiation-reaction force $\delta {\bm f}_{GR}$
is given by \cite{fll15}
\bea
&&
\!\!\!
\delta {\bm f}_{GR}=\sum_{l\geq 2}\sum_{|m|\leq l} \frac{(-1)^{\ell+1}N_\ell}{32\pi}
\,\Re\Biggl\{
\frac{\bm \nabla(r^\ell Y^{\ell m})}{\sqrt{\ell}}
\frac{d^{\,2\ell+1}\delta I^{\ell m}}{dt^{\,2\ell+1}}
\nonumber\\
&&
-
\frac{2r^\ell\bm Y^{\ell m}_{B}}{\sqrt{\ell+1}}
\frac{d^{\,2\ell+2}\delta S^{\ell m}}{dt^{\,2\ell+2}}
-
\frac{2\Omega\bm\phi
\times \bm \nabla(r^\ell Y^{\ell m})}{\sqrt{\ell}}
\frac{d^{\,2\ell+1}\delta S^{\ell m}}{dt^{\,2\ell+1}}\nonumber\\
&&
-
\frac{2\delta{\bm v}
\times \bm \nabla(r^\ell Y^{\ell m})}{\sqrt{\ell}}
\frac{d^{\,2\ell+1}\delta S^{\ell m}}{dt^{\,2\ell+1}}
\Biggr\}\,,
\label{e:PerturbedGRForceExact}
\eea
where
%
\bsube
\bea
\delta I^{\ell m}&:=&
\frac{N_\ell}{\sqrt{\ell}}
\int \delta \rho\, r^\ell Y^{*\ell m} d^3 x\,,
\label{e:PerturbedMassMultipole}\\
\delta S^{\ell m}&:=&\frac{2N_\ell}{\sqrt{\ell+1}}
\int r^\ell \left[\rho\,\delta {\bm v} +\delta \rho \,
\left(\Omega\bm\phi
+\delta {\bm v} \right)\right]\cdot \bm Y^{*\ell m}_{B} d^3x\,.\nonumber\\
\label{e:PerturbedCurrentMultipole}
\eea\esube

\subsection{Second-order axisymmetric perturbation}
\label{s:2ndorder}

The second-order perturbation of the MHD-Euler equation (\ref{e:FullEulerEquation})
has the form
\bea
\DeltaII \mathpzc E_i &=& (\partial_t+\Lie_{\bm v} )\DeltaII v_i + \nabla_i \DeltaII (U-\frac12 v^2)\nonumber\\
&&    + \frac1{2\pi\rho}B^j(\nabla_{[i}\DeltaII B_{j]}) = \DeltaII f_{GR\, i}\,.
\label{e:Delta2euler1}\eea
Here we have again used the commutation relation ~\eqref{e:DeltaLcommute} together with 
the commutator (also derived in Appendix~\ref{s:LagrangianPerturbations})
\be
\Delta d = d \Delta,
\label{e:commute1}\ee
where $d$ is the exterior derivative operator.

Equations~(\ref{e:Delta_rho}-\ref{e:Delta_B_i}) display the second-order
perturbation of each variable as a sum of two parts.  One part is linear
  in the second-order Lagrangian displacement $\xiII^i$,
  while the second part is quadratic in
  the first-order displacement $\xiI^i$. Each quantity is 
a sum of these two types of terms:
\be
   \DeltaII Q = \DeltaII_{\rm lin} Q + \DeltaII_{\rm quad} Q\,.
\ee
The linear part, $\DeltaII_{\rm lin} Q$
is the linear perturbation of $Q$ associated with the displacement
$\xiII^i$:  That is, $\DeltaII_{\rm lin} Q$ is identical to $\DeltaI Q$
if one replaces $\xiI^i$ by $\xiII^i$. This is essentially the statement
that, in the Taylor expansion of a function $F$ of $\xi^i$,
\bea
&&
\!\!\!\!\!
F(\alpha\xiI + \alpha^2\xiII) = F(0)+ \left.\frac{\partial F}{\partial
    \xi^j}\right|_{\xi=0} (\alpha\xiI^j +\alpha^2\xiII^j)
  \nonumber\\
  &&\qquad\qquad+
  \frac12\alpha^2 \left.\frac{\partial^2 F}{\partial
    \xi^j\pa\xi^k}\right|_{\xi=0} \xiI^j\xiI^k+O(\alpha^3)\,,
  \quad\nonumber
  \eea
$\alpha\xiI^i$ and $\alpha^2\xiII^i$ have the same coefficient, namely
the first derivative of $F$.

It follows that the second-order perturbation of the MHD-Euler equation
is again the sum of a part linear in the second-order Lagrangian
displacement $\xiII^i$ and a part quadratic in the first-order
displacement $\xiI^i$; similarly, $\DeltaII_{\rm lin}\mathpzc E_i$ is the
linear perturbation $\DeltaI \mathpzc E_i$ of Eq.~(\ref{e:DeltaIE_i2}), with
$\xiI^i$ replaced by $\xiII^i$. Including the second-order
radiation-reaction term, the second-order equation (\ref{e:Delta2euler1})
thus has the form
\bea
\rho\DeltaII \mathpzc E_i &=& A_{ij} \partial_t^2 \xiII^j + B_{ij} \partial_t\xiII^j +  C_{ij}  \xiII^j
\nonumber\\
&&    +D_i(\xiI,\xiI) = \rho\DeltaII f_{GR\, i}\,,
\label{e:Delta2euler2}
\eea
where the operators $A_{ij} $, $B_{ij} $, and $C_{ij} $ are given by Eqs.~(\ref{e:ABC})
and the quadratic operator $D_i$ has the form
\bea
&&
\!\!\!\!\!
\rho^{-1}D_i(\xiI,\xiI) = (\partial_t+\Lie_{\bm v})\DeltaII_{\rm quad} v_i
\nonumber\\
&&
\qquad\quad
+ \na_i \DeltaII_{\rm quad}\left(h+\Phi-\frac12 v^2\right)
            - \DeltaII_{\rm quad} f_{m\,i}\,.\qquad
\label{e:D_i}\eea
Here, with $\DeltaII_{\rm quad} B_i$ and $\DeltaII_{\rm quad} v_i$
displayed in Eqs.~(\ref{e:Delta_B_i}) and (\ref{e:Delta_v_i}), we obtain
\bsube
\bea
\DeltaII_{\rm quad} h
  &=&\frac12\frac{\gamma p}\rho
     \biggl[
       \left(\gamma-1 + \frac{\partial\log\gamma}{\partial\log\rho}\right)(\bm\na\cdot{\bm\xi}^{(1)})^2
       \nonumber\\
&&\qquad\quad        -\bm\xiI\cdot\bm\nabla\bm\na\cdot{\bm\xi}^{(1)}
     \biggr]\,,
\\
\DeltaII_{\rm quad} \Phi &=& \deltaII_{\rm quad}\Phi
+\bm\xiI\cdot\bm\nabla\deltaI\Phi
\nonumber\\
&&\quad                +\frac12\bm\xiI\cdot\bm\nabla(\bm\xiI\cdot\bm\nabla\Phi)\,,
\label{e:quadPhi}\\
  \DeltaII_{\rm quad}(\frac12 v^2)
      &=& \frac12\Bigl\{ \pa_t\xiI^i\pa_t\xiI_i
  +v^i2\pa_t \xiI^j\na_i\xiI_j
    \nonumber\\
    && \!\!\!\!\!\!\!\!\!\!\!\!\!\!\!\!\!\!\!\!\!
+v^i\pa_t(\xiI^j\na_j\xiI^i)+v^iv^j\Bigl(\xiI^k\na_k\na_i\xiI_j 
         \nonumber\\
         &&\!\!\!\!\!\!\!\!\!\!\!\!\!\!\!\!\!\!\!\!\!
         + \na_j\xiI^k\na_k\xiI_i
                            +\na_i\xiI^k\na_j\xiI_k \Bigr)
           \Bigr\}\,,
\\
  \DeltaII_{\rm quad}f_{m\,i}
      &=&  \frac1{2\pi\rho}B^j\nabla_{[j}\DeltaII_{\rm quad} B_{i]}\,.
\eea\esube
In Eq.~(\ref{e:quadPhi}) $\deltaI\Phi$ and $\deltaII_{\rm quad}\Phi$ are the potentials
associated with $\deltaI\rho$ and with $\deltaII_{\rm quad}\rho$:
\bsube\bea
  \na^2\deltaI\Phi &=& 4\pi\deltaI\rho = - 4\pi\na_i(\rho\xiI^i)\,,
\\
     \na^2\deltaII_{\rm quad}\Phi &=& 4\pi \deltaII_{\rm quad}\rho\,,
\\
   \deltaII_{\rm quad}\rho
   &=& \frac12\rho\left[(\na_i\xiI^i)^2 + \xiI^i\na_i \na_j\xiI^j\right]
   \nonumber\\
   &&\!\!\!\!\!\!\!\!\!\!\!\!\!\!
   +\xiI^i \na_i\rho\na_j\xiI^j + \frac12\xiI^i\na_i(\xiI^j\na_j\rho)\,,
\quad\qquad
\eea\esube
where the last expression is obtained from Eqs.~(\ref{e:DeltaIIQ}) and (\ref{e:Delta_rho}).


We now restrict consideration to an axisymmetric background star.
Because the components of $\xiI^i$ have time dependence $\cos(m\phi
+\omega t)e^{\beta t}$ and \mbox{$\sin(m\phi+\omega t)e^{\beta t}$} (see \cite{fll15}
and Eq.~\eqref{e:deltavN}), the
quadratic combination $D^i(\xiI,\xiI)$ is a sum of terms of three kinds:
terms with angular and temporal dependence $\cos[2(m\phi +\omega
  t)]e^{2\beta t}$, terms with dependence \mbox{$\sin[2(m\phi +\omega
    t)]e^{2\beta t}$}, and terms independent of $\phi$, with time
dependence $e^{2\beta t}$.

With the term $D_i(\xiI, \xiI)$ moved to
its right side, Eq.~(\ref{e:Delta2euler2}) has the form
\be
   A_{ij}  \pa_t^2\xiII^j + B_{ij} \pa_t\xiII^j + C_{ij} \xiII^j = \DeltaII F_i\,,
\label{e:inhom}\ee
where
\be
   \DeltaII F_i = \rho\DeltaII f_{GR\,i} - D_i(\xiI, \xiI)\,.
\ee

Recalling that we use brackets $\langle \cdot \rangle$ to denote the
axisymmetric part of a perturbation, we can write the axisymmetric part
of the second-order MHD-Euler equation as
\bea
  \langle\rho\DeltaII \mathpzc E_i\rangle &=& A_{ij} \partial_t^2 \langle\xiII^j\rangle
                    + B_{ij} \partial_t\langle\xiII^j\rangle
                    + C_{ij}  \langle\xiII^j\rangle
                    \nonumber\\
                &=& \langle \DeltaII F_i\rangle\,.
\label{e:Delta2euleraxisymm}\eea
Axisymmetry of the background star implies axisymmetry of the operators
$A_{ij}, B_{ij}$, and $C_{ij} $, allowing us to move the operators
outside the brackets. Acting on axisymmetric perturbations, the operator
$B_{ij} $ has the form
\bea
 B_{ij}  = -2\rho\epsilon_{ijk}\Omega^k\,,
\label{e:Bab1}\eea
where $\Omega^i$ is the angular velocity vector.
With the first-order perturbation $\xiI^i$ known, Eq.~(\ref{e:Delta2euleraxisymm})
is the equation for an axisymmetric linear perturbation of the star with a forcing term
\be
   \langle \DeltaII F_i \rangle := \rho\langle\DeltaII f_{GR\,i}\rangle
                    -\langle D_i(\xiI,\xiI)\rangle\,.
\label{e:DeltaIIF}\ee

   At second order in the perturbation, the star loses angular momentum to gravitational
waves. We can decompose the second-order axisymmetric perturbation into two parts,
one representing the spin down of the star, the other conserving total angular momentum.
The first part, $\deltaII_{UR} Q$, is a perturbation that adds {\it uniform rotation}
$\deltaII_{UR}\Omega<0$ to the star and has total (negative) angular momentum
equal to the angular momentum lost in gravitational waves; the second part, $\deltaII_{DR}Q$,
is the remaining, angular-momentum-conserving part of the second-order axisymmetric
perturbation that describes the addition of {\it differential rotation}
with zero total angular momentum. We write the corresponding decomposition of the
Lagrangian displacement in the form
\be
   \langle \xiII^i\rangle = \xi^{(2)\,i}_{UR} + \xi^{(2)\,i}_{DR}\,.
\ee
Finally, we can decompose the effective driving force $\langle\DeltaII F_i\rangle$
into an angular-momentum-reducing part that drives the change in uniform rotation
and an angular-momentum-conserving part.
\be
  \langle\DeltaII F_i\rangle = \DeltaII_{UR} F_i + \DeltaII_{DR} F_i\,,
\label{e:fur_fdr}\ee
where
\be
   \DeltaII_{UR} F_i := (A_{ij} \partial_t^2 + B_{ij} \partial_t+ C_{ij}  )\xi^{(2)\,b}_{UR}\,.
\label{e:fur}\ee

\subsection{Symplectic product and the growth of driven modes.}
\label{s:symplectic_summary}

We need an equation for the growth of the displacement $\xiII^i$ with a driving force
and a background magnetic field. The simplicity of the toy model comes from
fact that Eq.~(\ref{e:eulerphi}) governing the homogeneous solutions has
the form
\be
   \partial_t^2 \xi^\phi + C \xi^\phi = 0\,,
\ee
with the operator $C$ self-adjoint. This allows one to write the solution to the
inhomogeneous equation as a sum (Eq.~\ref{e:enorm}) of orthogonal eigenfunctions of
the operator $C$; and in the exponentially growing solution, the coefficient of each
eigenfunction of $C$ is proportional to the inner product of $f$ with the
normalized eigenfunction. In contrast, the dynamical equation (\ref{e:inhom}) governing
the $r$-mode includes a first time-derivative term with an operator $B^i{}_j$
that is anti-self-adjoint and does not commute with the self-adjoint operator $C^i{}_j$.
If that first-time derivative were not present, solutions to the homogeneous equation
could again be written as a superposition of eigenfunctions of $C^i{}_j$ and eigenfunctions
$\xi_n^i$ and $\xi_n'^i$ with distinct eigenvalues would be orthogonal with respect to the inner product
$\int dV  \xi^*_i\ A\eta^i = \int dV \rho  \xi^*_i\ \eta^i $.
The presence of the first-time derivative term means that solutions to the
homogeneous equation,
\be
  (A_{ij} \partial_t^2 + B_{ij} \partial_t + C_{ij} ) \xi^j =0\,,
\label{e:homogeneous}\ee
are not orthogonal in this sense. There is nevertheless
a conserved {\it symplectic} product with respect to which modes of the homogeneous
equation with distinct eigenvalues are orthogonal. We summarize the results
here and relegate to Appendix \ref{s:symplectic} a detailed derivation based
on Refs.~\cite{fs78,DS79} and a summary by Schenk {\it et al}. \cite{schenk02}.

Following Friedman and Schutz \cite{fs78}, we define the symplectic product of two complex solutions to the homogeneous equation
\mbox{$A\pa_t^2\xi + B\pa_t\xi + C\xi =0$}
by
\be
  W(\xi,\tilde\xi):= \inner\xi{\tilde\pi} - \inner\pi{ \tilde\xi}\,,
\label{e:symplectic}\ee
where $\pi_i$ is the momentum conjugate to $\xi^i$,
\be
\pi_i = \rho\pa_t\xi_i + \frac12 B_{ij} \xi^j\,,
\label{e:pi_i}\ee
and $\inner{\phantom{x}}{\phantom{x}}$ is the usual inner product
\be
  \inner\xi\eta = \int dV \xi^*_i\ \eta^i\,.
\ee
We use boldface angle brackets to distinguish the symbol for inner product from the ordinary
typeface brackets in the expression $\langle Q\rangle$ for the axisymmetric part of $Q$.

We will restrict consideration to perturbations that conserve total
angular momentum, mass and entropy; in particular, we use only the part
$\langle\Delta^{(2)}_{DR}F_i\rangle$ of the driving force in the decomposition
\eqref{e:fur_fdr} because the addition
of uniform rotation does not enhance the magnetic field.
We also assume that the linear axisymmetric modes of the axisymmetric background star
with magnetic field are stable, discrete and nondegenerate.  Because the operators
$A$, $B$ and $C$ are real, if $\xi$ satisfies the homogeneous equation so
does $\xi^*$. For a stable system with a complete set of discrete normal
modes, the modes therefore come in pairs
\be
\xi_n(t,x) = \widetilde\xi_n(x) e^{i\omega_n t}\,, \qquad \xi_n^* = \widetilde\xi_n^*(x) e^{-i\omega_n t}\,,
\label{e:xi_n}\ee
and we will write frequencies as $\pm\omega_n$, with $\omega_n>0$.
Because we are assuming a stable Newtonian system, the frequencies are real.
The fact that $W$ is conserved implies that modes with different frequencies are
symplectically orthogonal:
\be
   W(\xi_n,\xi_{n'}) = 0\,, \ \omega_n\neq \omega_{n'}, \qquad W(\xi_n^*,\xi_{n}) = 0\,.
\label{e:W_orthog}\ee

The proof is immediate: If $W(\xi_n,\xi_{n'})$ does not vanish, it has time dependence
$e^{i(\omega_{n'}-\omega_n)t}$, contradicting  $dW/dt = 0$.

With our assumption that the spectrum has no continuous part, work by Dyson and
Schutz \cite{DS79}, using symplectic orthogonality, shows that the modes are
complete. For a driving term of the form
\be
   F_i(t,x) = \hat F_i(x) e^{2\beta t},
\ee
their work implies [see our Appendix {\ref{s:symplectic},
    Eq.~(\ref{a:sumxin})] that the exponentially growing solution to the
  inhomogeneous equation,
\be
   (A_{ij} \partial_t^2 + B_{ij} \partial_t + C_{ij} ) \xi^j = F_i(t,x)\,,
\ee
is
\be
\xi^i = \sum_n \Re\left[\frac1{i\kappa_n\omega_n (2\beta - i\omega_n)}
              \inner{\widehat\xi_n}{F} \widehat\xi_n^i
                \right]\,,
\label{e:sumxin0}\ee
where the modes $\widehat\xi_n$ are normalized by
\be
  \inner{ \widehat\xi_n}{ \rho \widehat\xi_n } = 1\,,
\label{e:rhonorm}\ee
and
\be
 \kappa_n
     := 1 - 2\frac\Omega\omega_n\, \Im\frac{\int dV \rho\widetilde\xi_n^{\varpi*}\widetilde\xi_n^{\hat\phi}}{\int dV\rho|\widetilde\xi_n|^2}\,.
\label{e:kappa_n}\ee
We have adopted the convention $\omega_n > 0$; taking the real part of
the bracketed expression in Eq.~(\ref{e:sumxin0}) accounts for modes with frequency $-\omega_n$.
After saturation, the driving force is constant, and the displacement oscillates about a constant
equilibrium value given by Eq.~(\ref{e:sumxin0}) with $\beta=0$ (Eq.~(\ref{a:sumxinf}) of
Appendix {\ref{s:symplectic}),
\be
\xi^i = \sum_n \Re\left[ \frac1{\kappa_n\omega_n^2}
        \inner{\widehat\xi_n}{F} \widehat\xi_n^i
      \right]\,,
\label{e:sumxinf}\ee
where $F$ is the value of the driving force at saturation.}{}

Note that the canonical energy of the nth normalized mode is \cite{fs78}
\bea
E_{c\,n} &=& \frac12 W(\partial_t\widehat\xi_n,\widehat\xi_n)
\nonumber\\
     &=& -\frac12 i\omega_n W(\widehat\xi_n,\widehat\xi_n)
     = \omega_n^2\kappa_n \inner{ \widehat\xi_n}{ \rho\widehat\xi_n}\,.
\eea
If the unperturbed star is strictly stable against axisymmetric perturbations
(having neither unstable nor zero-frequency axisymmetric perturbations that conserve
angular momentum, baryon mass, and entropy),
then $E_{c\,n}>0$, implying $\kappa_n > 0$.

Finally, we break $\widehat\xi_n^i$ into its real and imaginary parts,
\be
  \widehat\xi_n^i = \widehat\xi_{nR}^i+i\widehat\xi_{nI}^i\,,
\ee
to elucidate the dependence of different contributions to the sum on
$\beta$, $\omega_n$ and $\Omega$.
A short calculation, beginning with the right side of Eq.~(\ref{e:sumxin0}) gives
\bea
  &&\xi^i
    =\sum_n \frac1{\kappa_n(4\beta^2+\omega_n^2)}
    \!\left[\inner{\widehat\xi_{nR}}{F}\widehat\xi_{nR}^i
        + \inner{\widehat\xi_{nI}}{F}\widehat\xi_{nI}^i\phantom{\frac12}\right.
\nonumber\\
       &&\qquad\qquad+ \left. \frac{2\beta}{\omega_n}
        \left(\inner{\widehat\xi_{nI}}{F}\widehat\xi_{nR}^i
        -\inner{\widehat\xi_{nR}}{F}\widehat\xi_{nI}^i\right)
    \right] \,.\qquad\quad
\label{e:cn_xiR_xiI}\eea
After saturation, Eq.~(\ref{e:sumxinf}) gives the equilibrium value
\be
  \xi^i
    =\sum_n \frac1{\kappa_n \omega_n^2}
    \left[\inner{\widehat\xi_{nR}}{F}\widehat\xi_{nR}^i
        + \inner{\widehat\xi_{nI}}{F}\widehat\xi_{nI}^i\right].
\label{e:xi_post_sat}\ee

\section{Growth of differential rotation and magnetic field windup}
\label{s:estimates}

To estimate the growth of the differential rotation of an unstable $r$-mode,
we use Eq.~(\ref{e:cn_xiR_xiI}) to write the solution
$\langle\xiII^\phi\rangle$ to Eq.~(\ref{e:Delta2euleraxisymm})
at saturation in the
form,
\bea
&&\!\!\!\!\!\!
\sum_n \frac1{\kappa_n(4\beta^2+\omega_n^2)}\times\nonumber\\
    &&\!\left[\inner{\widehat\xi_{nR}}{\langle\DeltaII F\rangle}\widehat\xi_{nR}^\phi
        + \inner{\widehat\xi_{nI}}{\langle\DeltaII F\rangle}\widehat\xi_{nI}^\phi\phantom{\frac12}\right.
\nonumber\\
     &&\quad+ \left. \frac{2\beta}{\omega_n}
    \left(\inner{\widehat\xi_{nR}}{\langle\DeltaII F\rangle}\widehat\xi_{nI}^\phi
      -\inner{\widehat\xi_{nI}}{\langle\DeltaII F\rangle}\widehat\xi_{nR}^\phi\right)
    \right]\,;\nonumber\\
\label{e:sumxin}\eea
after saturation, we use Eq.~(\ref{e:xi_post_sat}) to write the equilibrium value
of $\langle\xiII^\phi\rangle$ in the form
\be
\sum_n \frac1{\kappa_n\omega_n^2}
    \left[\inner{\widehat\xi_{nR}}{\langle\DeltaII F\rangle}\widehat\xi_{nR}^\phi
        + \inner{\widehat\xi_{nI}}{\langle\DeltaII F\rangle}\widehat\xi_{nI}^\phi\right].
\label{e:xipost}\ee
We estimate the value of the inner product
$\inner{\widehat\xi_n}{\DeltaII F}$ for modes $\xi_n$ whose $B=0$ limits
are zero-frequency axisymmetric perturbations associated with
differential rotation.  Primary differences between Eq.~(\ref{e:sumxin})
for the Lagrangian perturbation of the stellar model and
Eq.~(\ref{e:enorm}) for the toy model are (1) the effective driving force
$\DeltaII F$ includes the nonlinear terms $D_i$ as well as the
radiation-reaction force, and (2) the coefficient of the mode expression
for the Lagrangian displacement has the factor $1/\kappa_n$.

Although Eq.~(\ref{e:sumxin}) involves a sum over all axisymmetric modes,
modes with wavelengths much smaller than $R$ should give negligible
contributions, because the characteristic length of $\rho \langle\DeltaII
F\rangle$ is of order $R$ for the $\ell=m=2$ $r$-mode. (For smooth vector
fields $\bm f$ and $\bm g$, the inner product $\inner{g}{f}$ falls off
exponentially as the wavelength of the Fourier components of $\bm g$
approach zero.)  Of the axisymmetric modes with wavelengths of order $R$,
the Alfv\'en modes have the lowest frequencies, with magnitudes for
normal and superconducting interiors given by Eqs.~(\ref{eq:omega_A_normal})
and (\ref{eq:omega_A_superfluid}).  In
particular, normal-fluid $g$-modes have frequencies of order the
Brunt-V\"ais\"al\"a frequency of about 150 Hz (see,
e.g.,\cite{rg92,lai99}), and a class of superfluid $g$-modes has higher
frequency \cite{kg14,dg16}; inertial modes have frequencies of order
$\Omega$, and the frequencies of $p$- and $f$-modes are much higher.
Because the coefficient of the mode sum is proportional to $\omega^{-2}$
for $\omega\gg \beta$, we assume that the estimate is dominated by modes
with frequencies of order $\omega_A$.


We will find that the inner products of these axisymmetric Alfv\'en-frequency
modes with the two terms,
$\rho\DeltaII f_{GR\,i}$ and $-D_i(\xiI,\xiI)$
that comprise $\DeltaII F_i$ are of order
\be
  \inner{\widehat\xi_n}{ \rho |\DeltaII {\bm f}_{GR}|}\widehat\xi_n^\phi
 \sim  \beta \Omega e^{2\beta t}\,,
\qquad
  \inner{\widehat\xi_n}{D}\widehat\xi_n^\phi
 \sim  \omega_A\Omega e^{2\beta t}\,;
\label{e:estimate0}
\ee
%
As in the toy model, we set an upper limit on the maximum angular displacement
by adopting a driving force whose growth stops instantaneously at $t=t_{\rm sat}$.
Setting $\beta=0$ and $t=t_{\rm sat}$ in Eq.~(\ref{e:estimate0}) gives the
equilibrium values reached after saturation.

The estimates \eqref{e:estimate0} then imply (for $\beta,\,\omega_A\ll\Omega$) a maximum
value of the angular displacement at saturation given by
\be
   \xi^\phi_{\rm sat} = \alpha^2_{\rm sat} \xiII{}^\phi
    \sim \alpha^2_{\rm sat}
    \frac{{\rm max}(\omega_A,\beta)\,\Omega}{4\beta^2+\omega_A^2}\,
\label{e:estimate_drift}
\ee
and a maximum value after saturation
\be
  \xi^\phi_{\rm max}
    \sim \alpha^2_{\rm sat} \frac{\max(\omega_A,\beta)\,\Omega}{\omega_A^2}\,.
\label{e:estimate_post}\ee
As in the toy model, a larger post-saturation value of the displacement that
arises when $\omega_A < \beta$ is mitigated by a larger critical magnetic field
needed to alter the linear $r$-mode:  That is  after saturation, the critical
magnetic field is given by Eq.~(\ref{e:Bcritpost}) instead of Eq.~(\ref{e:B_kritical}).

We first outline the main ingredients that enter the estimates (\ref{e:estimate_drift}) and (\ref{e:estimate_post}) and then show how they are obtained.  We assume the linear $r$-mode grows exponentially
until a time $t_{\rm sat}$ and subsequently has constant amplitude.

\begin{itemize}

\item Prior to and at saturation, the radiation-reaction force per unit mass,
$\DeltaII f_{GR\, i}$
is of order
\be
 |\DeltaII {\bm f}_{GR}| \sim \beta\,\Omega\,R\,e^{2\beta t_{\rm sat}}\,.
\label{e:estimate_f}\ee
This immediately gives the first estimate in Eq.~(\ref{e:estimate0}).
\item With no magnetic field the quadratic contribution $D_i(\xiI_N,\xiI_N)$ from the
linear Newtonian $r$-mode $\xi_N^i$ has no $\phi$ component. With a generic magnetic
field, $\langle D_{\hat\phi}\rangle$ is small compared to $\langle D_\varpi\rangle$ and
$\langle D_z\rangle$:

\bea
  \rho^{-1}\left|\langle D_\varpi\rangle\right|
  &\sim& \rho^{-1}\left|\langle D_z\rangle\right|\sim \Omega^2\,R\, e^{2\beta t_{\rm sat}}\,,\nonumber\\
  \rho^{-1}\left|\langle D_{\hat\phi}\rangle\right|
    &\sim& \max(\omega_A^2, \beta\,\Omega)\, R\, e^{2\beta t_{\rm sat}}\,.
\label{e:estimate_D}
\eea

\item For the first-order axisymmetric modes $\xi^i_n$ associated with differential rotation,
the part of $\xi_n^i$ orthogonal to $\phi^i$ is small compared to $\xi_n^{\hat\phi}$:
\be
   |\xi^\varpi_n|,|\xi^z_n| \sim \frac{\omega_A}\Omega |\xi^{\hat\phi}_n|\,.
\label{e:estimate_xi}
\ee

This comes from the fact that, with no magnetic field, a perturbation
associated with adding differential rotation has the form $\DeltaI v^i =
\pa_t \xiI^i$, along $\phi^i$; Eq.~(\ref{e:estimate_xi}) estimates the
nonzero values of the components of $\bm\xi$ orthogonal to $\bm\phi$ for
a magnetic field with $\omega_A\ll \Omega$.
\item A consequence of the relations (\ref{e:estimate_xi}) is that the
  ratio of integrals that appears in the definition~(\ref{e:kappa_n}) of
  $\kappa_n$ has an upper bound of order
\[
   \frac{\left|\int dV \rho\widetilde\xi_n^{\varpi*}\widetilde\xi_n^{\hat\phi}\right|}
    {\int dV\rho|\widetilde\xi_n|^2} \lesssim \frac{\omega_A}{\Omega}\,,
\]
and this in turn gives an upper bound of order unity on $\kappa_n$,
\be
   |\kappa_n| \lesssim 1\,.
\label{e:estimate_kappa}\ee
\end{itemize}
The estimates~(\ref{e:estimate_D}) and (\ref{e:estimate_xi}) imply that the
quantity $\inner{\widehat\xi_n}{D}\langle\xi_n^{\hat\phi}\rangle$ has an
upper bound of order $\omega_A\, \Omega$, giving the second estimate in
Eq.~(\ref{e:estimate0}).  Finally, using the
estimate~(\ref{e:estimate_kappa}) for $\kappa_n$, we obtain our main
result, Eq.~(\ref{e:estimate_drift}).

To obtain the estimates (\ref{e:estimate_f}) and (\ref{e:estimate_D}) for
the two contributions to the effective driving force $\DeltaII
F=|\DeltaII \bm F|$, we will use the slow-rotation forms of the 
radiation-reaction force and the first-order Lagrangian displacement.  Corrections
are of order $\Omega/\Omega_0$, where $\Omega_0 =\sqrt{M/R^3}$. We use
the slow-rotation forms not because the corrections are negligible -- for
nascent stars with angular velocities near the Keplerian (mass-shedding)
limit $\Omega_K$, they could change the quantities we consider by factors
of order unity -- but because these corrections do not alter our
order-of-magnitude estimates.  We also neglect corrections to the linear $r$-mode
and radiation-reaction force due to the background magnetic field;
here the corrections {\it are} negligible for fields weaker than $10^{14}$-$10^{15}$ G
\cite{MR02,R02,lee05,GA07,LJP10,CS13,ARR12,aly15,Chugunov2015}.

We consider first the second-order radiation-reaction force,
$\langle\DeltaII f_{GR}^i\rangle$.  Because the radiation-reaction force vanishes
for the background star, Eq.~(\ref{e:DeltaIIQ}) gives as its second-order Lagrangian change
\be
   \DeltaII f_{GR}^i = \deltaII f_{GR}^i+\Lie_{\bm\xi^{(1)}} \deltaI f_{GR}^i\,.
\label{e:DeltaIIf}\ee
For the $\ell=m$ angular harmonic, the axisymmetric part of $\deltaII f_{GR}^i$
is given by [see Paper I, Eq.~(112)]
\be
 \bigl\langle \deltaIIRR\! f_{GR}^i\bigr\rangle
    = -\frac{(\ell+1)^2}4 \beta \Omega \left(\frac\varpi R\right)^{2\ell-2}e^{2\beta t}\phi^i\,,
\label{e:fGRSimplified}
\ee
at leading order in the star's angular velocity. The first-order
radiation-reaction force $\deltaI f_{GR}^i$ appearing in
Eq.~(\ref{e:DeltaIIf}) has the form
\be
  \deltaI f_{GR}^i = \beta\deltaI v^i + \deltaI_\perp f_{GR}^i\,,
\label{e:deltaIf}\ee
where $\langle\bm \xi^{(1)}, \deltaI_\perp {\bm f}_{GR}\rangle = 0$ [see
  Paper I, Eq.~(86)]. Because of this orthogonality, $\beta\deltaI v^i$,
determines the growth rate of the linear mode $\xiI^i$.

 At leading order in $\Omega$, $\delta v^i$ and $\xi^i$ are orthogonal to $\widehat{\bm r}$,
and their components along unit vectors $\hat\theta$ and $\hat\phi$ are
\bsube\bea
 \deltaI v^{\hat\theta}
 &=& \widetilde\deltaI v^{\hat\theta} \cos(\ell\phi + \omega t)e^{\beta t}
 \nonumber\\
 &=& -\Omega R \left(\frac rR\right)^\ell\sin^{\ell-1}\theta
                \cos(\ell\phi + \omega t)e^{\beta t}\,, \qquad \\
\deltaI v^{\hat\phi}
&=& \widetilde\deltaI v^{\hat\phi} \sin(\ell\phi + \omega t)e^{\beta t}
\nonumber\\
&=& \Omega R \left(\frac rR\right)^\ell\sin^{\ell-1}\theta\cos\theta
                \sin(\ell\phi + \omega t)e^{\beta t}\,,\nonumber\\
\eea\label{e:deltavN}\esube
\bsube\bea
\xi^{(1)\,\hat\theta} &=& \widetilde\xi^{\hat\theta} \sin(\ell\phi + \omega t)e^{\beta t}
\nonumber\\
&=& -\frac{\Omega}{\omega_r} R
                 \left(\frac rR\right)^\ell\sin^{\ell-1}\theta
                   \sin(\ell\phi + \omega t)e^{\beta t}\,,  \\
                   \xi^{(1)\,\hat\phi} &=& \widetilde\xi^{\hat\phi} \cos(\ell\phi + \omega t)e^{\beta t}
                   \nonumber\\
            &=& -\frac{\Omega}{\omega_r} R
               \left(\frac rR\right)^\ell\sin^{\ell-2}\theta\cos\theta
                \cos(\ell\phi + \omega t)e^{\beta t}\,,
\nonumber\\
                \eea\label{e:xiN}\esube
where, to leading order in $\Omega$, $\displaystyle\omega =
-\frac{(\ell-1)(\ell+2)}{\ell+1}\Omega$ and $\displaystyle\omega_r =
\frac2{\ell+1}\Omega$ is the frequency in a rotating frame.  From
Eqs.~(\ref{e:deltavN}) and (\ref{e:xiN}), the vectors $\xiI^i$ and
$\deltaI v^i$ are of order
\be
    \xiI\sim R e^{\beta t}, \qquad \deltaI v \sim \Omega R e^{\beta t}\,.
\label{e:deltav_xi_est}
\ee
The divergence $\bm\na\cdot{\bm\xi}^{(1)}$ vanishes at lowest order in
$\Omega$, and is nonzero only at order $\Omega^2$ \cite{pbr81}, with
\be
\bm\na\cdot{\bm\xi}^{(1)} \sim \frac{\Omega^2}{\Omega_0^2}e^{\beta t}\,,
\label{e:divxi}\ee
where
\be
  \Omega_0 := \sqrt{\frac{GM}{R^3} }\sim \frac{v_s}{R}\,,
\ee
with $v_s$ an average speed of sound in the star.

Prior to saturation, from Eqs.~(\ref{e:fGRSimplified}) and~(\ref{e:deltaIf}),  $\deltaII f^i$ and
$\deltaI f^i$ are of order $\beta\,\Omega\,R\,e^{2\beta t}$ and $\beta\,\Omega\,R\,e^{\beta t}$,
respectively. Then Eq.~(\ref{e:deltav_xi_est}) implies
the term $\Lie_{\bm\xi^{(1)}}\deltaI {\bm f}_{GR}$ is of order
\[
   \left|\Lie_{\bm\xi^{(1)}} \deltaI {\bm f}_{GR}
        \right|\sim \beta\,\Omega\,R\,e^{2\beta t}\,,
\]
and we obtain the estimate~(\ref{e:estimate_f}), $|\DeltaII {\bm f}_{GR}|
\sim \beta\,\Omega\,R\,e^{2\beta t}$.

We turn next to Eq.~(\ref{e:estimate_D}) for $\langle
D_i(\xiI,\xiI)\rangle$, where $\xiI$ is the Lagrangian displacement of
the first-order unstable $r$-mode. To estimate $\langle D_i\rangle$, we
use Eqs.~(\ref{e:deltav_xi_est}) and (\ref{e:divxi}), together with the
estimate $\nabla Q \sim Q/R$. From Eq.~(\ref{e:divxi}), we have
\be
\frac{\DeltaI\rho}{\rho} \sim \frac{\DeltaI p}p
            \sim \frac{\Omega^2}{\Omega_0^2}\, e^{\beta t}\,.
\ee
Equation (\ref{e:D_i}) gives $\langle D_i(\xiI, \xiI)\rangle$ as a
sum of three terms which we consider in order. The angle average removes
both the $\phi$ dependence and the harmonic dependence on $t$, leaving
only the dependence $e^{2\beta t}$. We then have
\be
  |(\partial_t+\Lie_{\bm v} )\langle\DeltaII v_i\rangle_{\rm quad}|
    = 2\beta | \langle\DeltaII v_i\rangle_{\rm quad}|
    \sim \beta\, \Omega\, R\, e^{2\beta t}.
\ee
The $\phi$ component of the second term on the right of Eq.~(\ref{e:D_i})
vanishes by axisymmetry: $\displaystyle\pa_\phi\langle U-\frac12 v^2\rangle = 0$;
the components orthogonal to $\phi^i$ have magnitudes of order
\bsube\bea
 |\nabla \langle\DeltaII_{\rm quad} U\rangle|
    &=& |\nabla \langle\DeltaII_{\rm quad}(h+\Phi)\rangle|
 \sim \Omega^2\, R\, e^{2\beta t}\,, \nonumber\\
 \\
 |\nabla \langle\DeltaII_{\rm quad}\frac12 v^2\rangle|
    &\sim& \Omega^2\,R\, e^{2\beta t}.
\eea\esube
The last, magnetic term of Eq.~(\ref{e:D_i}) is of order
\be
   \left|\frac1{2\pi\rho}B^j(\nabla_{[i}\langle\DeltaII_{\rm quad} B_{j]}\rangle)\right|
    \sim \omega_A^2\,R\, e^{2\beta t}\,.
\ee
From Eq.~\eqref{e:t_A}, the Alfv\'en frequency for a type II superconductor has the form
\be
 \!\!\!  \omega_{A,SC} = \frac1R\sqrt{ \frac{\pi B_0 H_c}{\rho}}
    = 0.09\  R_6^{-1}\,\sqrt{ \frac{B_{9}\, H_{c,15}}{\rho_{14.6}}}
   \ {\rm s^{-1}} \,.
\label{eq:omega_A_superfluid}
\ee
This and Eq.~\eqref{e:omega_A} for a normal fluid each imply $\omega_A < \Omega$ unless
$B_0 > 10^{17}$ G.  Then for both
nascent neutron stars and old accreting neutron stars, rotating fast enough
to be unstable to an $r$-mode, we have $\omega_A\ll \Omega$, and we recover Eq.~(\ref{e:estimate_D}),
\bsube\bea
  \rho^{-1}\left|\langle D_\varpi(\xiI, \xiI)\rangle\right|
    &\sim& \rho^{-1}\left|\langle D_z(\xiI, \xiI)\rangle\right|
     \nonumber\\&\sim& \Omega^2\,R\, e^{2\beta t}\,, \\
\rho^{-1}\left|\langle D_{\hat\phi}(\xiI, \xiI)\rangle\right| 
    &\sim& \max(\omega_A^2, \beta\,\Omega)\, R\, e^{2\beta t}\,.
\qquad
\eea\esube

Finally, we justify the estimate~(\ref{e:estimate_xi}). That is, we show
that $\xi^\varpi$ and $\xi^z$ are of order
$(\omega_A/\Omega)\,\xi^{\hat\phi}$ for an axisymmetric solution $\xi^i$
to the perturbed MHD-Euler equation whose $B=0$ limit is a perturbation
that describes a change in the rotation law -- the addition of
differential rotation to a uniformly rotating star. Like the vanishing of
$D_\phi$, the estimate is related to the form of the Euler equation for
axisymmetric perturbations.  Writing $\mathpzc E_i$ for a general fluid with no
magnetic field in the form
\be
  \mathpzc E_i = (\partial_t + \Lie_{\bm v})v_i + \frac{\nabla_i p}\rho +\na_i(\Phi -\frac12 v^2)\,,
\ee

we have
\be
  \mathpzc E_\phi = (\partial_t +\Lie_{\bm v})v_\phi\,,
\ee
with $\mathpzc E_\phi=0$ expressing angular momentum conservation of each fluid ring.
The commutator in Eq.~(\ref{e:DeltaLcommute}) implies
\be
  \Delta \mathpzc E_\phi = \partial_t \Delta v_\phi\,.
\ee

The fact that only the time-derivative term survives means, for a
first-order axisymmetric perturbation described by a Lagrangian
displacement $\xiI^i$
\be
  \phi^i \DeltaI \mathpzc E_i = \phi^i\pa_t \DeltaI v_i =\phi^i( \pa_t^2 \xiI_i +2 \epsilon_{ijk}\Omega^j\partial_t\xiI^k)\,,
\ee
implying that the operator $C_{ij} $ has no component along $\phi^i$. When a background
magnetic field is present, $C_{ij} $ acquires a nonzero $\phi$ component given
by the last line on the right of Eq.~(\ref{e:Cab}), with magnitude
\be
 \rho^{-1} C_{\hat\phi j}\xiI^j \sim \frac{B^2\xiI}{\rho R^2}\sim \omega_A^2\, Re^{\beta t}\,.
\label{e:C_phi}
\ee

The corresponding magnitude of $\langle\xiII{}^\varpi\rangle$ can be seen
from the $\hat\phi$ component of the second-order Newtonian Euler equation:
\begin{align}
 \partial_t^2 \langle\xiII_{\hat\phi}\rangle + 2\Omega \partial_t \langle \xi^{(2)\varpi}&\rangle
  + \rho^{-1} C_{\hat\phi\, j}\langle \xi^{(2)j}\rangle 
\nonumber\\
  &= - \rho^{-1} \langle D_{N\hat\phi}(\xiI,\xiI)\rangle\,.
\end{align}
The first-order axisymmetric modes satisfy
\be
  \partial_t^2 \xi_{n\hat\phi} + 2\Omega \partial_t \xi_n^\varpi
    + \rho^{-1} C_{\hat\phi\, j}\xi_n^j = 0\,.
\ee
We approximate the frequencies of the dominant modes by $\omega_A$,
  writing $\partial_t \xi_n \sim \omega_A\xi_n$, $\partial_t^2 \xi_n\sim
  \omega_A^2\,\xi_n$, and use Eq.~(\ref{e:C_phi}) to write $\rho^{-1}
  C_{\hat\phi\, j}\xi_n^j \sim \omega_A^2\, \xi_n$. We then have
\be
    \xi_n^\varpi \sim \left(\frac{\omega_A}{\Omega}\right) \xi_n^{\hat\phi}\,.
\ee
Finally, in the expression~(\ref{e:kappa_n}) for $\kappa_n$,
\[
 \kappa_n
     = 1 - 2
\left(\frac\Omega\omega_n\right)
\Im\frac{\int dV \rho\,\widetilde\xi_n^{\varpi*}\widetilde\xi_n^{\hat\phi}}{\int dV\rho\,|\widetilde\xi_n|^2}\,,
\]
the ratio of integrals is of order $\omega_A/\Omega$, giving a bound on
$\kappa_n$ of order unity.  This completes our justification of the
estimates (\ref{e:estimate_D}), (\ref{e:estimate_xi}), and
(\ref{e:estimate_kappa}); and the argument following
Eq.~(\ref{e:estimate_kappa}) then gives our main result,
Eq.~(\ref{e:estimate_drift}) for the angular displacement of a fluid
element.

\noindent{\it Normal interior}

We turn now to the implications of this estimate.
We first find bounds on magnetic
field growth for a normal interior and then obtain equivalent
bounds for an interior that is a type II superconductor.
We obtain as follows a bound on the maximum growth of
$\delta B$ similar to Eq.~\eqref{e:B_toy_limit} of the toy model.
In Eq.~\eqref{e:estimate_drift},
\[
    \frac{{\rm max}(\omega_A,\beta)}{4\beta^2+\omega_A^2}
  = \max\left[\frac{\omega_A}{4\beta^2+\omega_A^2},\ \frac{\beta}{4\beta^2+\omega_A^2}\right].
\]
By inspection, $\displaystyle \frac{\omega_A}{4\beta^2+\omega_A^2} < \frac1{\omega_A}$,
and, using the inequality \eqref{e:ineq}, we have
\[
   \frac{{\rm max}(\omega_A,\beta)}{4\beta^2+\omega_A^2} < \frac1{\omega_A}.
\]
Then the angular displacement and corresponding change in the magnetic field
have upper limits
\be
\langle\xi^\phi_{\rm sat}\rangle \lesssim \alpha_{\rm sat}^2\frac\Omega{\omega_A}
< \alpha_{\rm sat}^2\frac{\Omega R}{B_0} \sqrt{\frac\rho\pi} ,
\label{e:xi_bound1}\ee
\be
  \langle\delta B^{\hat\phi}_{\rm sat}\rangle \lesssim \alpha_{\rm sat}^2 \Omega R \sqrt{\frac\rho\pi},
\label{e:B_bound1}\ee
with the small numerical values
\bea
  \langle\xi^\phi_{\rm sat}\rangle &\lesssim& 0.4\ \alpha_{-4}^2 f_{500}R_6 B_9^{-1}\rho_{14.6}^{1/2}, \nonumber\\
 \langle\delta B^{\hat\phi}_{\rm sat}\rangle &\lesssim& 4\times 10^8\ \alpha_{-4}^2 f_{500}R_6 \rho_{14.6}^{1/2}\ \rm G.
\eea
Recalling Eq.~(\ref{e:B_kritical}) for the critical magnetic field and using Eq.~\eqref{e:B_bound1},
we obtain our main inequality,
\be
  \frac{\langle \delta B_{\rm sat}^{\hat\phi}\rangle}{\langle\delta B\rangle_{\rm crit}}
    \lesssim \alpha_{\rm sat} ,
\label{e:B_bound2}\ee
or, equivalently,
\be
  \frac{d{\cal E}_m
/dt} {d{\cal E}_{\rm mode}/dt}
    \lesssim \alpha_{\rm sat}^2.
\ee

When $\alpha_{\rm sat} \sim \mathcal{O}(1)$, as assumed in the
  initial investigations of the instability \cite{Owen1998} and in
  Refs.~\cite{Rezzolla00, Rezzolla01b, Rezzolla01c}, then $\langle \delta
  B_{\rm sat}^{\hat\phi}\rangle \sim \langle\delta B\rangle_{\rm crit}$,
  and the magnetic field at saturation is similar to the critical field
  needed to damp or substantially alter the linear $r$-mode. However, for
  more realistic values of the saturation amplitude, and even for an
  unexpectedly large saturation amplitude, $\alpha_{\rm sat}\sim
  10^{-3}$, the change in the magnetic field at saturation is three
  orders of magnitude below the critical field.

After nonlinear saturation, the constraint on $\langle \delta
B^{\hat\phi}\rangle$ corresponding to the limit (\ref{e:estimate_post})
on the angular displacement is \be \langle \delta B^{\hat\phi}_{\rm max}
\rangle \lesssim \alpha_{\rm sat}^2 B_0
        \begin{cases}
               \Omega/{\omega_A}, & \omega_A > \beta \\
                       {\beta\Omega}/{\omega_A^2}, & \omega_A < \beta .
    \end{cases}
\ee
With the critical magnetic field now given by Eq.~(\ref{e:Bcritpost}),
\begin{align}
\langle\delta B\rangle_{\rm crit} &\sim \alpha_{\rm sat} \Omega R\sqrt{\rho\beta/\omega_A}
     \geq  \frac{\alpha_{\rm sat}}{\pi^{1/4}} \beta^{1/2}\Omega R^{3/2}\rho^{3/4}B_0^{-1/2}, 
\end{align}
we have
\bsube
\bea
\frac{\langle \delta B^{\hat\phi}_{\rm max}\rangle} 
     {\langle\delta B\rangle_{\rm crit} }
&\lesssim&
     \alpha_{\rm sat}\sqrt{\displaystyle\frac{\ \omega_A}{\!\!\pi\beta}}
      \qquad\qquad\qquad\mathrm{for} \quad\omega_A > \beta
\nonumber\\
&\leq& 2.4\times10^{-5} \alpha_{-4}\beta_{-3.3}^{-1/2}R_6^{-1/2}\rho_{14.6}^{-1/4} B_9^{1/2}, 
         \nonumber\\\\ 
\frac{\langle \delta B^{\hat\phi}_{\rm max}\rangle}{\langle\delta B\rangle_{\rm crit} }
    &\lesssim&
\alpha_{\rm sat}\sqrt{\displaystyle\frac{\beta}{\!\pi\omega_A}}
      \qquad\qquad\qquad\mathrm{for} \quad\omega_A < \beta
\nonumber\\
        &\leq& 1.3\times 10^{-4}
            \alpha_{-4}\beta_{-3.3}^{1/2} R_6^{1/2}\rho_{14.6}^{1/4}B_9^{-1/2} 
                 .\nonumber\\
            \label{e:B_post_bound}\eea\esube
The second case ($\omega_A < \beta$) is Eq.~\eqref{e:Bpost_limit} of the toy model. For $\omega_A > \beta$, the present bound differs
from that of the toy model because of the contribution
to the effective driving force from the quadratic $D$ term, but not by enough to
alter our conclusion.

  In particular, after saturation, the oscillation may allow $\xi_{DR}^{(2)\,\phi}$
to grow to about twice its equilibrium value, with a smaller value for a more
gradual approach to saturation.
Even with $\alpha_{\rm sat} \sim 10^{-3}$, the initial magnetic
field would need to be well below 100 G or above $10^{16}$ G before magnetic field windup
could significantly alter the linear $r$-mode.

\noindent{\em Superconducting interior}

The $r$-mode instability has been studied most in the context of old neutron stars
spun up by accretion. The interior of these stars is likely to be a type II superconductor,
and we now turn to the corresponding limits on magnetic-field windup
for  such stars.

For a superconducting interior, the total energy of the magnetic field is given by
\be
   E_{m,SC} =  \frac{1}{8\pi} \varphi_{\rm SC}\, H_c\, \ell_f \,,
\ee
where $\ell_f$ is the average length of a flux tube, and
$\varphi_{\rm SC}$ is the total magnetic flux.
Differential rotation stretches the flux tubes but leaves
the flux in each tube and the number of tubes unchanged.
Then $\varphi_{\rm SC}$ is constant, and the change in energy
$E_{m,SC}$ is determined by the change in flux tube length $\ell_f$.
For a tube deformed by a small angular displacement $\langle\xi^\phi\rangle$,
the change in length at quadratic order in $\xi^\phi$ is of order
\begin{equation}
\delta \ell_f\approx \ell_f\langle\xi^\phi\rangle^2\,.
\end{equation}
With $\ell_f\sim R$, the stretching rate at quadratic order is then
\begin{equation}
\frac{d \ell_f}{dt}\sim R \xi^{\phi}\frac{d\langle\xi^\phi\rangle}{dt}\,
            =2\beta R (\xi^{\phi})^2,
\end{equation}
We define a field $B_0$ for which the total flux is
\be
   \varphi_{\rm SC} = \pi R^2 B_0.
\label{e:BSC}\ee
The total magnetic energy is then
\be
  E_{m,SC} = \frac18 B_0 H_c \ell_f R^2,
\ee
larger than its value for a normal plasma by a factor of order
$H_c/B_0$,
and the corresponding growth rate of magnetic energy density is
\begin{equation}
\frac{d {\cal E}_{m,SC}}{dt}\sim \frac1{30}\beta H_c B_0 (\xi^{\phi})^2,
\label{dESC}
\end{equation}
for a superconducting core of approximate radius $R$.
A detailed calculation
by Rezzolla {\it et al.}~\cite{Rezzolla00, Rezzolla01b, Rezzolla01c} for an initial dipole poloidal magnetic field $B_0$
gives the same relation with a somewhat smaller numerical coefficient,
\be
 \frac{d{\cal E}_{m,SC}}{dt}
    \sim \beta \frac{1}{60} B_0 H_c\, \langle\xi^\phi\rangle^2.
\label{e:EmSC}\ee

We define an average perturbed magnetic field, $\langle\delta B_{\rm SC}\rangle$, as
a volume average for which $\langle\delta B_{\rm SC}\rangle^2/8\pi := \delta {\cal E}_m $.
The critical magnetic field for which the growth rate of magnetic energy and
of the linear $r$-mode are equal is then again given by Eq.~(\ref{e:B_kritical}).

To obtain an approximate bound on $d{\cal E}_m/dt$ and $\langle\delta B_{\rm SC}\rangle$,
we first write Eq.~\eqref{e:EmSC} in the form
\be
  \frac{d{\cal E}_{m,SC}}{dt}
    \sim \beta \frac{1}{60\pi} \rho\omega_{A,SC}^2 (R\langle\xi^\phi\rangle)^2.
\label{e:EomegaA}\ee
The bound on $\langle\xi^\phi\rangle$ is given by Eq.~\eqref{e:xi_bound1}
with $\omega_A$ replaced by $\omega_{A,SC}$,
\be
  \langle\xi^\phi_{\rm sat}\rangle \lesssim \alpha_{\rm sat}^2\frac\Omega{\omega_{A,SC}}
< \alpha_{\rm sat}^2\Omega R \sqrt{\frac\rho{\pi B_0 H_c}},
\label{e:xi_boundSC}
\ee
with the small numerical value
\be
  \langle\xi^\phi_{\rm sat}\rangle
    \lesssim 6\times 10^{-4} \alpha_{-4}^2 f_{500} \sqrt{\frac{\rho_{14.6}}{B_9 H_{c,15}}}.
\ee
We then have
\be
  \frac{d{\cal E}_{m,SC}}{dt}
    \lesssim \frac{1}{60\pi} \alpha_{\rm sat}^4 \beta  \rho\Omega^2 R^2.
\label{e:dmSC1}\ee
Recognizing that the right side is proportional to the energy of the linear
$r$-mode, as in Eq.~\eqref{eq:E_mode}, we obtain the inequalities
\begin{equation}
 \frac{\langle \delta B_{{\rm sat},SC}\rangle}{\langle \delta B_{SC}\rangle_{\rm crit}} \lesssim \frac1{\sqrt{60\pi}} \alpha_{\rm
   sat}\,,\qquad \frac{d{\cal E}_{m,SC}/dt} {d{\cal E}_{\rm mode}/dt}
 \lesssim \frac{1}{60\pi} \alpha_{\rm sat}^2\,.
\end{equation}

We are not entitled to claim bounds this stringent, however, because in deriving
the bound $\langle\xi^\phi\rangle\lesssim \alpha_{\rm sat}^2\Omega/\omega_{A,SC}$,
 we used the rough approximation
$\omega_n \sim \omega_{A,SC}$, while the coefficient $1/60\pi$ in the
expression for $d{\cal E}_m/dt$ is consistent with the somewhat smaller frequency
of long-wavelength Alfv\'en modes. What our estimates show are then
the approximate bounds previewed in Sec.~\ref{s:magnitudes},
\be
 \frac{\langle \delta B_{\rm SC}\rangle_{\rm sat}}{\langle \delta B_{\rm SC}\rangle_{\rm crit}}
    < \alpha_{\rm sat}\,,\qquad
   \frac{d{\cal E}_{m,SC}/dt} {d{\cal E}_{\rm mode}/dt}
    \lesssim \alpha_{\rm sat}^2.
\ee

After saturation, because $\omega_{A,SC} \gg \beta$, the maximum displacement
and magnetic field are within a factor of about 2 of their values at saturation.

\noindent{\it Caveats: Continuous spectrum, zero-frequency modes, and MRI instability}

The claim that magnetic field windup cannot damp or significantly alter the
first-order $r$-mode comes with some caveats.  The estimates of this
section rely on two principal assumptions: That linear axisymmetric
perturbations of the background star can be written in terms of a
discrete nondegenerate spectrum, and that the background star has no
unstable axisymmetric modes - or at least no unstable axisymmetric modes
that wind up the magnetic field.

It may be that neither assumption is correct: There is no proof that discrete
modes are complete for uniformly rotating stars, and, once differential rotation
is established, the star is likely to encounter a magnetorotational instability (MRI).
We briefly discuss the implications of relaxing the assumptions, beginning with
a possible continuous part of the spectrum of linear modes.

Because the effective driving force $\DeltaII \bm F$ is a quadratic function of
the linear $r$-mode, its value is unrelated to assumptions about the
spectrum of linear axisymmetric perturbations. With a continuous spectrum,
estimates~(\ref{e:estimate_f}) and (\ref{e:estimate_D}) of its two parts
are unchanged, and $\DeltaII \bm F$ retains its form, with magnitude
\be
 \DeltaII F_{\hat\phi} \sim \max(\omega_A^2,\beta\,\Omega)\, R\, e^{2\beta t}.
\ee
Were we able to replace a sum over discrete modes by an integral over a
continuous spectrum, we could regain our estimates for $\xi^\phi$. We
have no formal justification for this, because the time evolution of the
system is described by an operator that is not self-adjoint. Simply
discretizing the spatial operators, however, gives a system whose modes
{\it are} discrete and for which 
the estimates hold. Because
the estimates are independent of the discretization, they should hold in
the continuum limit.

The assumption of a stable system is in question once differential
rotation is established by a growing $r$-mode. That is, there appears to
be an MRI instability when the magnetic field is smaller than about
$10^{13}$ G \cite{chandra61,bh94} and the drift angular velocity
$\delta\Omega_{drift}$ satisfies
\be
   \frac{d(\delta\Omega^2_{drift})}{d\varpi} < 0\,,
\ee
in some region of the star. The instability is present only for
perturbations that are not restored by negative buoyancy or by
pressure. Buoyancy is governed by the Brunt-V\"ais\"al\"a frequency,
which, for a neutron star, is of order 50-150 Hz (see, e.g.,
\cite{rg92,lai99}), much larger than $R\ d\Omega/d\varpi \sim \alpha(t)^2
\Omega$. In the Balbus-Hawley analysis \cite{bh94}, this removes the
instability for most modes, but leaves at least a set of unstable
perturbations whose wavevector $\bm k$ is quasiradial, along the
Brunt-V\"ais\"al\"a vector $\bm N$, and there may also be modes with zero
or near zero frequency. Because MRI-unstable perturbations cannot acquire
more energy than is present in the small available differential rotation,
we suspect that the presence of MRI-unstable or marginally unstable
perturbations will not substantially alter our analysis. We should point
out, however, that after saturation, the constant effective
radiation reaction force $\DeltaII F^a$ will drive the growth of any
zero-frequency modes.

\section{Conclusions}
\label{sec:discussion}

Almost 20 years ago, $r$-mode oscillations in rotating neutron stars
were shown to be unstable to the emission of gravitational waves
\cite{A97,FM98}. The impact of this finding on newly born neutron stars
and in old neutron stars in X-ray binaries was soon discussed in a long
list of works starting with Ref. \cite{Owen1998}. Among the many features
of the nonlinear development of the stability, the development of
differential rotation was pointed out early on, heuristically \cite{Spruit99}
and via perturbation theory \cite{Rezzolla00}, as was the
amplification of strong magnetic fields and the possibility that this
growth suppresses the instability \cite{Rezzolla01b,Rezzolla01c}.

Building on a more realistic estimate of the saturation amplitude of
  the instability \cite{Bondarescu07,Bondarescu09} and on a more rigorous
  mathematical description of the development of differential rotation in
  unstable stars \cite{fll15}, we have here reconsidered the impact of
  differential rotation and magnetic field
amplification on the growth of unstable $r$-modes. The instability may be
present in nascent neutron stars and in old stars in X-ray binaries;
in each case, nonlinear coupling to other modes limits the $r$-mode amplitude
to a saturation amplitude $\alpha_{\rm sat} \lesssim 10^{-4}$. And in each case
we find that the maximum enhancement of the average magnetic field is
smaller by the factor $\alpha_{\rm sat}$ than the critical field needed
to damp or significantly alter the $r$-mode.
 We have obtained this result following two different
  routes: First, using a simplified but exact toy model where the star is
  treated as an incompressible and homogeneous cylinder in the ideal-MHD
  limit; second, using a formalism governing the equilibrium and first-
  and second-order perturbations of a rotating star with a background
  magnetic field and radiation reaction.

  In old neutron stars whose interior is a type II superconductor, we
  find that magnetic-field growth stops soon after the mode reaches its
  saturation amplitude.  In nascent neutron stars, before the interior
  has cooled below the superconducting transition temperature, continued
  magnetic-field growth can follow nonlinear saturation.  If the
  saturation amplitude is unexpectedly large, with
  $\alpha_{\rm sat}\sim 10^{-3}$, an initial small
  magnetic field of about $10^8$ G could be amplified to $10^{11}$ G,
  before the remaining secular drift of a fluid element (that winds up
  the magnetic field) is restricted to less than a radian. Although still
  too small to damp the growth of the linear $r$-mode, this might be a
  contribution to magnetic-field generation in nascent stars.

Although mathematically robust, our findings rest on the assumptions
noted at the end of the last section. In particular, we assume
that there are no marginally unstable perturbations, and this may
not hold when differential rotation leads to a magnetorotational instability.

\bibstyle{prd}
\bibliography{./References}

\acknowledgments

\noindent We thank Ruxandra Bondarescu, Mikhail E. Gusakov, Stuart Shapiro, Branson
Stephens, and Ira Wasserman for helpful conversations.  LL was supported
in part by NSF grants PHY 1604244 and DMS 1620366 to the University of
California at San Diego. LR was supported in part by ''NewCompStar'',
COST Action MP1304, from the LOEWE-Program in HIC for FAIR, the European
Union's Horizon 2020 Research and Innovation Programme under grant
agreement No. 671698 (call FETHPC-1-2014, project ExaHyPE), from the ERC
Synergy Grant ``BlackHoleCam - Imaging the Event Horizon of Black Holes''
(Grant 610058), and from JSPS Grant-in-Aid for Scientific Research(C)
No. 26400274. The work by A. I. C. consisted in supporting consideration of B-field
amplification (supported by the Russian Science Foundation, grant No. 14-12-00316).


\appendix

\section{Lagrangian perturbations}
\label{s:LagrangianPerturbations}

At first order in $\alpha$ the Lagrangian displacement vector $\bm\xi =
\alpha\bm\xiI$ can be viewed in two ways.  $\bm\xi$ is a connecting
vector from the position $x$ of a fluid element in the unperturbed fluid
to its position $\chi_\alpha(x)$ in the perturbed fluid; and $\bm\xiI$ is
the vector field tangent to the trajectories $\alpha\rightarrow
\chi_\alpha(x)$ of the family of diffeomorphisms $\chi_\alpha$.  At
higher order the two viewpoints diverge and we have chosen the second
approach, defining a Lagrangian displacement that depends only on the
family of diffeomorphisms, not on the metric of flat space or on a
choice of coordinates.  The second-order formalism using the first approach
is developed in Ref.~\cite{fs78}.

\subsection{First- and second-order Lagrangian perturbations}
\label{a:2nd_order_displacement}

    We derive here relations used in Sec.~\ref{s:perturbations} to
obtain first- and second-order Lagrangian perturbations, defined by
Eq.~(\ref{e:LagrangianDelta2}).

Recall that the pullback map $\chi^*$ associated with a diffeomorphism $\chi$
is defined on scalars $f$ by
\be
\chi^* f(t,x) := f(t,\chi[t,x])\,.
\ee
On covariant and contravariant vectors $w_i$ and $w^i$ its action
is given in any coordinate system by
\bsube
\bea
\chi^* w_i(t,x) &=& \partial_i \chi^j\, w_j(t,\chi[t,x])\,,
\\
\chi^* w^i(t,x) &=& \partial_j\left( \chi^{-1}\right)^i
w^j(t,\chi[t,x])\,.
\eea\esube
Acting on forms (antisymmetric covariant tensors) $\omega_{a\ldots b}$, it satisfies
\be
   [\chi^*, d] \omega = 0\,,
\label{e:chi*d}\ee
where $d$ is the exterior derivative.

 Given a family of diffeomorphisms $\chi_\alpha(x)$ of the unperturbed
 fluid to the perturbed fluid at a fixed time $t$, we can define a
 family of Lagrangian displacements $\xi(\alpha,x)$ in a way that is
 analogous to defining the velocity field $v^i(t,x)$ from the family
 of diffeomorphisms $\psi_t$ that describe the fluid flow: In the
   fluid case the family of diffeomorphisms acts on both the
   spatial coordinates $x$ and the time coordinate $t$, while in our
   analogous case the parameter $\alpha$ plays the same role as the
   time coordinate in the fluid case.  In the time-dependent fluid
   case $\psi_\tau$ maps a fluid element at $x$ at a time $t$ to its
 position $\psi_\tau(t,x)$ at time $t+\tau$. The velocity field
 $v^i(t,x)$ is tangent to the curve $c(\tau) = \psi_\tau(t,x)$.  \be
 v^i(t,x) = \left.\frac d{d\tau} c^i(\tau)\right|_{\tau=0} =
 \left.\frac d{d\tau} \psi^i_\tau(t,x)\right|_{\tau=0}\,.  \ee More
 concisely, the four-dimensional diffeomorphism $\Psi_\tau$, \be
 \Psi_\tau(t,x) = (t+\tau, \psi_\tau(t,x))\,,
\label{a:Psi_tau}\ee
moves the point $(t,x)$ a parameter distance $\tau$ along an integral curve of the
Newtonian 4-velocity
\be
   {\bm u}(t,x) = (1,v^i(t,x))\,.
\ee

We now repeat the construction for the family of diffeomorphisms
$\chi_\alpha(x)$.  In this case, we include the parameter $\alpha$ as a
coordinate and denote by $(\alpha,x)$ a point in the support of the perturbed
fluid: The fluid element at $(0,x)$ in the unperturbed fluid is at the
corresponding point $(\alpha, \chi_\alpha(x))$ in the perturbed fluid.
As initially defined, $\chi_\alpha$ maps a point $x$ occupied by a fluid
element in the unperturbed fluid to the location $\chi_\alpha(x)$ of that
fluid element in the perturbed fluid.  We extend $\chi_\alpha$ to a
family $\widetilde\chi_\alpha$ of diffeomorphisms that act on points in
the perturbed fluid by writing
\be
   \widetilde\chi_\eta(\alpha,\chi_\alpha(x)) := \chi_{\eta+\alpha}(x)\,.
\label{a:chi_tilde}\ee
We define the vector field $\widetilde\bxi(\alpha,x)$ as the tangent to the curve
$c(\eta) = \widetilde\chi_\eta(\alpha,x)$,
\be
   \widetilde\xi^i(\alpha,x) = \left.\frac d{d\eta} c^i(\eta)\right|_{\eta=0}
        = \left.\frac d{d\eta} \widetilde\chi^i_\eta(\alpha,x)\right|_{\eta=0}\,,
\label{a:xi_tilde}\ee
to maintain a Lagrangian displacement $\bxi$ that is proportional to $\alpha$ at
lowest order, we write
\be
   \bxi = \alpha\widetilde\bxi\,.
\label{a:xi_def}\ee
Again our construction has a more concise form in terms of the four-dimensional diffeomorphism
$X_\eta$ (the analog of $\Psi_\tau$),
\be
   X_\eta(\alpha,x) = (\alpha+\eta, \widetilde\chi_\eta(\alpha,x))\,:
\ee
$X_\eta$ moves the point $(\alpha,x)$ a parameter distance $\eta$ along an integral curve of the
vector field
\be
   {\bm\Xi}(\alpha,x) = (1,\widetilde\xi^i(\alpha,x))\,.
\ee

This is the statement that $\bm\Xi$ generates the family of diffeomorphisms  $X_\alpha$,
and it leads to a simple expression, (\ref{a:DeltaQ}) below, for the Lagrangian
perturbation in the fluid variables $Q$ at nth order in $\alpha$.
We begin by noting that the relation
\be
   \frac d{d\alpha} X_\alpha^* f(x) = \frac d{d\alpha}  f(X_\alpha(x))
        = \left.(\Lie_\bXi f)\right|_{X_\alpha(x)}\,,
\ee
for a scalar $f$, implies
\be
  \left.\frac {d^n}{d\alpha^n}  f(X_\alpha(x))\right|_{\alpha=0} = \Lie_{\Xi}^n f(x)\,.
\ee
The action of an analytic family of diffeomorphisms  $X_\alpha$ on an analytic function
is then given by a convergent Taylor series in $\alpha$, namely
\be
  X_\alpha^* f = e^{\alpha\Lie_{\mathbf \Xi}} f\,.
\ee
In our case, we have only a smooth family of diffeomorphisms  acting on a smooth function, and
the Taylor series at finite order in $\alpha$ gives the relation
\be
  X_\alpha^* f = \left[1+ \alpha\Lie_{\mathbf \Xi} + \cdots + \frac1{n!}(\alpha\Lie_{\mathbf \Xi})^n
        + o(\alpha^n)\right]f\,.
\ee
It is straightforward to check that the same relation holds for the action of $X_\alpha^*$
on arbitrary smooth tensors.

From the definition ~(\ref{e:LagrangianDelta1}) of the exact Lagrangian change in the fluid variables $Q(\alpha,x)$, we have
\be
   X_\alpha^* Q(0,x) = Q(\alpha,\chi_\alpha(x))\,,
\ee
implying
\be
   \Delta Q = X_\alpha^* Q(0,x) - Q(0,x)
    =  \sum_1^n \alpha^k\frac1{k!}\left.\Lie_{\mathbf \Xi}^k Q\right|_{\alpha=0} + o(\alpha^n)\,.
\label{a:DeltaQ}\ee
In particular, writing
\bsube
\bea
   \bxiI &=& \left.\pa_\alpha\bm\xi\right|_{\alpha=0}
    = \left.\widetilde{\bm\xi}\right|_{\alpha=0},\\
   \bxiII &=& \left.\frac12 \pa_\alpha^2\bm\xi\right|_{\alpha=0}
    = \left.\pa_\alpha\widetilde{\bm\xi}\right|_{\alpha=0},
    \eea
    \esube
and $\bm\Xi_0 := \bm\Xi|_{\alpha=0}$, we obtain
\bea
\Delta^{(1)} Q &=& \left.\Lie_\bXi Q\right|_{\alpha=0}
\nonumber\\
    &=& \left.(\partial_\alpha+\Lie_{\tilde\bxi}) Q\right|_{\alpha=0}
    = (\deltaI + \Lie_{\widetilde\bxi^{(1)}}) Q\,,\\
 \Delta^{(2)} Q
    &=& \left( \Lie_{\pa_\alpha\bm\Xi} Q
               + \frac12\left.\Lie_{\bm\Xi_0}^2 Q
        \phantom{\frac12}\!\right)\right|_{\alpha=0}\nonumber\\
     &=& \left[ \Lie_{\pa_\alpha\widetilde{\bm\xi}}
             + \frac12\left.(\partial_\alpha + \Lie_{\widetilde{\bm\xi}^{(1)}})^2 Q
       \phantom{\frac12}\!\right] \right|_{\alpha=0}
\nonumber\\
        &=&\left.\left( \frac12\pa_\alpha^2 + \Lie_{\pa_\alpha\widetilde{\bm\xi}}
            + \Lie_{\widetilde{\bm\xi}^{(1)}}\pa_\alpha + \frac12\Lie_{\widetilde{\bm\xi}^{(1)}}^2
             \right) Q \right|_{\alpha=0}
\nonumber\\
        &=& \left( \deltaII + \Lie_{\bm\xi^{(2)}} + \Lie_{\bm\xi^{(1)}}\deltaI
                + \frac12\Lie_{\bm\xi^{(1)}}^2
            \right) Q\,.
\label{e:DeltaIDeltaII}\eea
In these last two equations, we have used the definition (\ref{e:deltaN Q Def}) of
$\delta^{(n)} Q$.

\subsection{Perturbed fluid velocity}
\label{s:Delta_v}

We will next find the expression for the Lagrangian change in the fluid velocity
in terms of the Lagrangian displacement of the fluid, obtaining the form
\be
\Delta v^i = \partial_t \xi^i + \half \Lie_{\bxi}\partial_t\xi^i
+{\cal O}(\alpha^3)\,.
\label{a:Delta_v}\ee
Expanding this result in powers of $\alpha$ immediately gives
\bea
\DeltaI v^i &=& \partial_t \xiI^i\,,\\
\DeltaII v^i &=& \partial_t \xiII^i + \half \Lie_{\bxiI}\partial_t\xiI^i\,.
\eea
Equation~(\ref{a:Delta_v}) can be derived by noting that the diffeomorphism
$\chi$ maps trajectories in the unperturbed fluid to trajectories in the perturbed fluid.
Denote by $\tau\mapsto c_0(t+\tau)$ the path of the fluid
element in the unperturbed fluid that passes through the point
$x=c_0(t)$ at time $t$. Then $\tau\mapsto \chi_\alpha(t+\tau,c_0(t+\tau))$
is the path of the fluid element in the perturbed flow, and it passes
through $\chi_\alpha(t,x)$ at time $t$. The perturbed velocity is then given by
\bea
   v_\alpha^i(t,\chi_\alpha(t,x)) &=& \left.\frac d{d\tau} \chi_\alpha^i(t+\tau,c_0(t+\tau))\right|_{\tau=0}\nonumber\\
   &=& \partial_t \chial^i+ v^k_0\partial_k\chial^i\,.
\eea
The exact Lagrangian change in the fluid velocity is given by
\bea
\Delta v^i(t,x)
&=& \chial^* v_\alpha^i(t,\chial(t,x)) - v_0^i(t,x)\,,\\
&=&
\left.\partial_j (\chial^{-1})^i\right|_{(t,\chial(t,x))}
v_\alpha^j(t,\chial(t,x))- v^i_0(t,x)\,.\nonumber\\
\label{a:Delta_vi}\eea
In all the remaining equations, each variable is evaluated at
the point $(t,x)$ unless the argument is explicitly shown.
Note first that, by its definition (\ref{a:chi_tilde}),
   $\widetilde\chi_\eta(\alpha,\chi_\alpha(x)) = \chi_{\eta+\alpha}(x)$.
From Eq.~(\ref{a:xi_tilde}), we then have
\bea
   \widetilde\xi^i(\alpha,\chi_\alpha(x))
        &=&\left.\frac d{d\eta}\chi_{\eta+\alpha}^i(x)\right|_{\eta=0}
   = \frac d{d\alpha}\chi_\alpha^i(x)\,,
   \nonumber\\
   \xi^{(1)i}(x) &=& \left.\frac d{d\alpha}\chi_\alpha^i(x)\right|_{\alpha=0}\,.
\eea
Similarly,
\bea
\left.\frac {d^2}{d\alpha^2}\chi_\alpha^i(x)\right|_{\alpha=0}
\!\!\!\!\!
        &=& \left.\frac d{d\alpha}\widetilde\xi^i(\alpha,\chi_\alpha(x))\right|_{\alpha=0}
   \nonumber\\
   &=& \left[\frac d{d\alpha}\widetilde\xi^i(\alpha,x)
            + \partial_j\widetilde\xi^i(0,x)\frac d{d\alpha}\chi_\alpha^i(x)
           \right]_{\alpha=0}
\nonumber\\
    &=& 2\xiII^i + \xiI^j\partial_j\xiI^i\,.
\eea
The expansion of the diffeomorphism $\chi_\alpha$,
\be
  \chial^i(x) = x^i
           + \alpha\pa_\alpha \chial^i\biggr|_{\alpha=0}
           +\frac12\alpha^2 \left.\pa_\alpha^2\chial^i\right|_{\alpha=0}
        + \mathcal{O}(\alpha^3)\,.
\label{e:DiffeoExpansion}
\ee
now gives
\bea \chial^i &=& x^i + \xi^i +
\half\xi^j\partial_j\xi^i + {\cal O}(\alpha^3)\,,\\
\chial^{-1\,i}&=& x^i - \xi^i +
\half\xi^j\partial_j\xi^i + {\cal O}(\alpha^3)\,.
\eea
Using these expressions, we obtain
\be
\left.\partial_j (\chial^{-1})^i\right|_{(t,\chial(t,x))}
= \delta^i{}_j - \partial_j\xi^i
-\xi^k\partial_k\partial_j \xi^i
+\half\partial_j\left(\xi^k\partial_k \xi^i\right)\,,
\label{e:Jacobian}
\ee
and
\bea
v_\alpha^j(t,\chial(t,x)) &=&
\partial_t \chial^j
+ v^k_0\partial_k\chial^j\,,\nonumber\\
&=& \partial_t \xi^j
+\half\partial_t\left(\xi^k \partial_k \xi^j\right)
+v_0^j
+v_0^k\partial_k\xi^j
\nonumber\\
&&+\half v_0^\ell\partial_\ell\left( \xi^k\partial_k\xi^j\right)
+{\cal O}(\alpha^3)\,.
\label{e:PerturbedVelocity}
\eea
Substituting in Eq.~(\ref{a:Delta_vi}) the expressions from Eqs.~(\ref{e:Jacobian}) and
(\ref{e:PerturbedVelocity}) and keeping terms up to quadratic order in $\bm \xi$ yields
the desired expression (\ref{a:Delta_v}) for $\Delta \bm v$.

\subsection{Commutation relations}

We now derive the commutation relations used in Sec.~\ref{s:2ndorder},
namely \footnote{
At first order, Eq.~(\ref{e:DeltaLcommute}) can be obtained by using the relation
\[
   \left[\Lie_{\bm \xi},\Lie_{\bm v}\right] = \Lie_{[\bm\xi, {\bm v} ]} \,,
\]
to write
\[
  \left[\DeltaI, (\partial_t + \Lie_{\bm v} )\right]
    = -\Lie_{\partial_t {\bm v} } + \Lie_{[\bm\xiI, {\bm v} ]}
    = \Lie_{-\partial_t {\bm v} +[\bm\xiI, {\bm v} ]} = 0\,.
\]
This algebraic derivation can be extended to the more complicated second-order commutator,
but it hides the simpler connection between the commutator (\ref{e:DeltaLcommute}) and
the commutation relation of the diffeomorphisms, Eq.~(\ref{e:diffeocommute}). }
\bsube\bea
   \Delta (\partial_t + \Lie_{\bm v} ) &=& (\partial_t + \Lie_{{\bm v} _0}) \Delta\,,
\label{e:DeltaLcommute}\\
\Delta d &=& d \Delta\,,
\label{e:dLcommute}\eea\label{e:app_kommute}\esube
where the second relation is restricted to an action on forms.

We first show that Eq.~(\ref{e:DeltaLcommute}) follows from a commutation relation between
the diffeomorphism $\chi_\alpha$ and the diffeomorphism generating the fluid flow. It is
simplest to write the relation in terms of the corresponding four-dimensional diffeomorphisms .
Let ${\cal X}_\alpha$ be the spacetime diffeomorphism associated with $\chi_\alpha$,
\be
  {\cal X}_\alpha(t,x) = (t,\chi_\alpha(t,x))\,,
\ee
and let
\be
  t\mapsto C_\alpha(t) = (t, c_\alpha(t))\,,
\ee
be the trajectory of a fluid element in the perturbed fluid, with Newtonian 4-velocity
$(1,{\bm v} )$, where $v^i(t) = \dot c_\alpha^i(t)$.
Then
\be
   C_\alpha(t) = {\cal X}_\alpha\comp C_0(t)\,.
\ee

As in Eq.~(\ref{a:Psi_tau}), let $\Psi_{\tau,\alpha}$ be the spacetime diffeomorphism that
maps a fluid element at time $t$ in the perturbed fluid to its position at time $t+\tau$:
\be
  \Psi_{\tau,\alpha} \comp C_\alpha(t) = C_\alpha(t+\tau)\,.
\ee
Then
\be
   \Psi_{\tau,\alpha}\comp {\cal X}_\alpha\comp C_0(t) = C_\alpha(t+\tau) = {\cal X}_\alpha\comp \Psi_{\tau,0}\comp C_0(t)\,,
\ee
implying
\be
   \Psi_{\tau,\alpha}\comp {\cal X}_\alpha = {\cal X}_\alpha\comp \Psi_{\tau,0}\,.
\label{e:diffeocommute}\ee
The Lie derivative of a tensor $T$ with respect to the 4-velocity $(1,{\bm v} )$
is
\be
(\partial_t + \Lie_{\bm v}) T = \left.\frac d{d\tau} \Psi_{\tau,\alpha}^*T\right|_{\tau=0}\,,
\ee
where $\Psi_{\tau,\alpha}^*$ is the pullback map. By Eq.~(\ref{e:diffeocommute}) the corresponding
pullbacks satisfy
\be
  {\cal X}_\alpha^* \Psi_{\tau,\alpha}^* = \Psi_{\tau,0}^*{\cal X}_\alpha^*\,.
\ee
Finally, taking the derivative of this relation with respect to $\tau$ at $\tau=0$, we obtain
Eq.~(\ref{e:DeltaLcommute}) for tensors $T$ that are functions of $\alpha$ and $x$:
\bea
 (\pa_t+\Lie_{{\bm v} _0})\Delta Q
&=& \left.\frac d{d\tau} \Psi_{\tau,0}^* ({\cal X}_\alpha^* Q_\alpha - Q_0)\right|_{\tau=0}
\nonumber\\
    &=& \left.\frac d{d\tau}( {\cal X}_\alpha^*\Psi_{\tau,\alpha}^* Q_\alpha-\Psi_{\tau,0}^* Q_0)\right|_{\tau=0}
\nonumber\\
    &=& \Delta(\partial_t+\Lie_{\bm v}) Q\,.
\eea

The second commutation relation, Eq.~(\ref{e:dLcommute}), is immediate from the
vanishing commutator of exterior derivative and pullback (acting on forms)
\be
   [d,\chi_\alpha^*] = 0\,.
\ee

\section{Symplectic product and the growth of driven modes}
\label{s:symplectic}

We derive here Eq.~(\ref{e:sumxin0}) for the growth of a system satisfying an
equation of the form
\be
   (A_{ij} \partial_t^2 + B_{ij} \partial_t + C_{ij} ) \xi^j = F_i(t,x)\,.
\label{a:inhom}
\ee
This is essentially a summary of results due to Dyson and Schutz
\cite{DS79}, included here because their work and the summary given by
Schenk {\it et al}. \cite{schenk02} are more elaborate, including in
particular the Jordan chains that arise when there are degenerate
modes. The treatment here is self-contained if one assumes that the
discrete normal modes are a complete set for arbitrary initial
data. Schutz and Dyson have a lengthy characterization of the spectrum
that implies completeness of the discrete modes if one assumes only that
the spectrum has no continuous part.

As noted Sec.~\ref{s:symplectic_summary}, orthogonality of nondegenerate
modes follows from the fact that the symplectic product $W$ of Eq.~(\ref{e:symplectic})
is conserved. This is a property of any Hamiltonian system. Here, a quick computation,
using only the self-adjointness properties of the operators, the homogeneous equation,
and the definition (\ref{e:pi_i}) of $\pi_i$, gives a direct check that
$d/dt\ W(\xi,\tilde\xi) = 0$.

For a nonrotating star, the quantity $i W(\xi_n,\xi_n)$ is real and is,
for each mode with nonzero frequency, proportional to the usual norm
$||\cdot||$, given by $||\xi||^2 = \inner\xi{A\xi} = \int dV \rho
|\xi|^2$.  Because the constant of proportionality involves $\omega_n$,
and, even for spherical stars, $i W(\xi_n,\xi_n)$ has no definite sign,
we will use $W$ itself to normalize $\xi_n$, writing
\bea
1 &=& W(\xi_n, \xi_n)\nonumber\\
&=& \inner{\xi_n}{A\pa_t\xi_n+\frac12 B \xi_n}
            - \bm\langle A\pa_t\xi_n+\frac12 B\left. \xi_n\right| \xi_n \bm\rangle\nonumber\\
  &=&  \inner{ \xi_n}{2i\omega_n A\xi_n + B \xi_n}\,.
\label{a:wxin}\eea
We now assume that the modes are nondegenerate,
\be
    \omega_n \neq \omega_{n'}\,, \qquad \mbox{for } n\neq n'\,,
\ee
implying the orthogonality relation (\ref{e:W_orthog})\,,
\[
  W(\xi_n,\xi_{n'}) = 0\,, \quad \omega_n\neq \omega_{n'}, \qquad W(\xi_n^*,\xi_{n}) = 0\,,
\]
and we assume that there are no zero-frequency modes. We adopt the convention $\omega_n>0$ and
write a general solution to the homogeneous equation in the form
\bea
\xi &=& \sum_n \left(C_{n+} \xi_n  +  C_{n-} \xi_n^*\right)
\nonumber\\
   & =& \sum_n\left( C_{n+} \widetilde \xi_n e^{i\omega_n t} +  C_{n-}\widetilde \xi_n^* e^{-i\omega_n t}\right)\,,
\eea
where $\xi_n(t,x) = \widetilde\xi_n(x) e^{i\omega_n t}$.
The coefficients $C_{n\pm}$ are then given by
\be
  C_{n+} = W(\xi_n,\xi)\,, \quad C_{n-} = W(\xi_n^*,\xi)\,.
\ee
For a real solution, we have $C_{n-} = C_{n+}^*$.

The familiarity of an expansion in terms of orthonormal eigenfunctions belies
a subtlety of the system:  Completeness of the modes means completeness of
the pairs of initial data
\be
(\xi_{n\pm}, \pa_t\xi_{n\pm})|_{t=0} = (\widetilde\xi_n, \pm i\omega_n \widetilde\xi_n)\,.
\ee
That is, arbitrary initial data $(\xi,\pa_t\xi)_{t=0}$ in the domain
of the operators has a spectral decomposition of the form
\be
\left.\begin{pmatrix}
  \xi \\
  \pa_t\xi
\end{pmatrix}\right|_{t=0}
=
C_{n+}\begin{pmatrix}
  \widetilde\xi_n \\
  i\omega_n\widetilde\xi_n
\end{pmatrix} +
C_{n-}\begin{pmatrix}
  \widetilde\xi_n^* \\
  -i\omega_n\widetilde\xi_n^*
\end{pmatrix}\,.
\label{a:Cn}
\ee
The coefficients $C_{n\pm}$ in the expansion of $\xi$ appear to determine
the coefficients $\pm i\omega_n C_{n\pm}$ in the expansion of $\partial_t\xi$. 
How is this possible, when
$\xi$ and $\pa_t\xi$ are each arbitrary?  The explanation is that the two sets
of eigenfunctions $\{\widetilde\xi_n\}$ and $\{\widetilde\xi_n^*\}$ are
not linearly independent; thus in Eq.~(\ref{a:Cn}) the equation for $\xi$
(or for $\partial_t\xi$) alone does not determine $C_{n+}$ and
$C_{n-}$. Each set $\{\widetilde\xi_n\}$ and $\{\widetilde\xi_n^*\}$ is
separately a basis for the configuration space $H$ of the system, and
using both gives a basis $\{(\widetilde\xi_n,
i\omega_n\widetilde\xi_n),(\widetilde\xi_n^*,
-i\omega_n\widetilde\xi_n^*)\}$ for the set $H\times H$ of pairs $(\xi,
\ \pa_t\xi)$.

This behavior -- the fact that the set $\{\widetilde\xi_n\}$ of vectors
associated with $\{\omega_n\}$ and the set $\{\widetilde\xi_n^*\}$ of
vectors associated with $\{-\omega_n\}$ are each a basis for $H$ is clear for
the homogeneous equation of a spherical star.  Here a mode satisfies
\be
   -\omega_n^2 A\xi_n + C\xi_n = 0\,.
\ee
If the eigenvalue $\omega_n^2$ is nondegenerate, then the normalized eigenvectors
associated with $\omega_n$ and $-\omega_n$ differ only by a constant phase; they
coincide as rays in a Hilbert space.
In the more general case of a stable rotating star with a discrete spectrum,
the fact that the sets $\{\xi_n\}$ and $\{\xi_n^*\}$
are each a basis for $H$ is shown by Dyson and Schutz.

Consider now a solution $\xi(t)$ to the inhomogeneous equation (\ref{e:inhom}).
Completeness of the normal modes for data on each constant $t$ hypersurface means
that, at each time $t$, we can find coefficients $c_{n\pm}(t)$ that satisfy
\be
\begin{pmatrix}
  \xi \\
  \pa_t\xi
\end{pmatrix}
= \sum_n\left[
c_{n+}(t)\begin{pmatrix}
  \widetilde\xi_n \\
  i\omega_n\widetilde\xi_n
\end{pmatrix} +
c_{n-}(t)\begin{pmatrix}
  \widetilde\xi_n^* \\
  -i\omega_n\widetilde\xi_n^*
\end{pmatrix} \right] \,.
\label{a:expansion}\ee
By inserting the eigenfunction
expansion into this equation and using the symplectic product
$W$ to project onto each mode $\widetilde\xi_n$, we will find for $c_{n\pm}(t)$
the dynamical equations
\bsube\bea
\dot c_{n+} - i\omega_n c_{n+} &=& \inner{ \widetilde\xi_ n}{F}\,, \\
  \dot c_{n-} + i\omega_n c_{n-} &=& \inner{ \widetilde\xi_ n^*}{F}\,.
\eea\esube

The derivation is as follows. From its definition (\ref{e:symplectic}), $W$ can be regarded
as acting on pairs $(\xi,\pa_t\xi)$ and $(\eta, \pa_t\eta)$ of data at a time $t$, with
\bea
&&  W[(\xi,\pa_t\xi);(\eta,\pa_t\eta)] := W(\xi,\eta) \nonumber\\
  &&
\qquad\quad  = \inner{ \xi}{ A\pa_t\eta+\frac12 B \eta}
            - \inner{ A\pa_t\xi+\frac12 B \xi}{ \eta }
\,.\qquad
\label{a:Wdef}
\eea
For mode data $(\widetilde\xi_n, i\omega_n\widetilde\xi_n)$, the relations $A^\dagger=A, B^\dagger = -B$ give
\bea
  W[(\widetilde\xi_n, i\omega_n\widetilde\xi_n);(\eta,\pa_t\eta)]
  &=& \inner{ \widetilde\xi_ n}{ A\pa_t\eta+i\omega_n A\eta + B \eta}\,,
  \nonumber\\
\eea
and Eq.~(\ref{a:expansion}) then implies
\bea
c_{n+}(t) &=& W[(\widetilde\xi_n, i\omega_n\widetilde\xi_n);(\xi(t),\pa_t\xi(t))]
\nonumber\\
    &=&  \inner{ \widetilde\xi_ n}{ A\pa_t\xi+i\omega_n A\xi + B \xi}\,.
\label{a:cn+}\eea
Taking the time derivative of this equation and using Eq.~(\ref{e:inhom})
to replace $A\pa_t^2\xi$ by \mbox{$-B\pa_t\xi - C\xi + F$}, we obtain
\bea
 \dot c_{n+}(t) &=& \inner{ -C\widetilde\xi_n}{ \xi}
    + \inner{ \widetilde\xi_ n}{ i\omega_n A\pa_t\xi_n + F}\,.
\label{a:cdot1}\eea
The homogeneous equation for the mode $\xi_n$ implies
\be
  C \widetilde\xi_n = \omega_n^2 A\widetilde\xi_n - i\omega_n B\widetilde\xi_n\,,
\ee
whence
\bea
  \inner{ -C\widetilde\xi_n}{ \xi}
  &=& \inner{ -\omega_n^2 A\widetilde\xi_n + i\omega_n B\widetilde\xi_n}{ \xi} \nonumber\\
&=& i\omega_n\inner{ \widetilde\xi_ n}{i\omega_n A\xi + B\xi}\,.
\label{a:Cxi}\eea
Finally, from Eqs.~(\ref{a:Cxi}) and (\ref{a:cdot1}), we have
\bea
  \dot c_{n+}(t)
    &=& i\omega_n \inner{ \widetilde\xi_n}{ A\pa_t\xi + i\omega_n A \xi + B\xi}
        + \inner{ \widetilde\xi_ n}{F}  \nonumber\\
    &=& i\omega_n c_{n+}(t) + \inner{ \widetilde\xi_ n}{F}\,,
\eea
with Eq.~(\ref{a:cn+}) used to obtain the last equality.
The same steps with $c_{n+},\ \xi_n$ and $\omega_n$ replaced by $c_{n-},\ \xi_n^*$ and $-\omega_n$, respectively, yield the corresponding equation for $\dot c_{n-}(t)$.
To summarize, the driven system is governed by the equations
\bsube\bea
\dot c_{n+} - i\omega_n c_{n+} &=& \inner{ \widetilde\xi_ n}{F}\, \\
  \dot c_{n-} + i\omega_n c_{n-} &=& \inner{ \widetilde\xi_ n^*}{F}\,.
\eea\label{a:cn}\esube

For an exponentially growing driving force $F_i(t,x) = \widetilde F_i(x) e^{2\beta t}$,
the mode amplitudes of the particular solution $\xi^i$
to Eq.~(\ref{a:inhom}) with time dependence $e^{2\beta t}$ are given by
\be
 c_{n+}(t) = c_{n-}^*(t)
    = \frac1{2\beta - i\omega_n} \inner{ \widetilde\xi_n}{\widetilde F} e^{2\beta t}\,,
\label{a:cnpm}\ee
and we have
\be
   \xi^i
    = \sum_n 2\Re\left[\frac1{2\beta-i\omega_n}
        \inner{ \widetilde\xi_n}{F}\widetilde\xi_n^i
                 \right]\,.
\label{a:xi_p}\ee

To estimate the magnitude of $\xiII$ in Sec.~\ref{s:estimates}, it is
helpful to rewrite this expression in terms of mode functions $\widehat\xi_n^i$
normalized by
\be
  \inner{ \widehat\xi_n}{ \rho \widehat\xi_n } = 1\,.
\label{a:rhonorm}\ee
We first find the symplectic norm of the mode functions $\widetilde\xi_n$.
From Eqs.~(\ref{e:Aab}), (\ref{e:Bab1}) and (\ref{a:wxin}), we have
\bea
1 &=& W(\widetilde\xi_n,\widetilde\xi_n)
\nonumber\\
&=& \inner{ \widetilde\xi_n}{ 2i\omega_n A\widetilde\xi_n + B \widetilde\xi_n}
\nonumber\\
    &=& \inner{ \widetilde\xi_{n\,i}}{ 2i\omega_n \rho\widetilde\xi_n^i
        -2\rho\Omega \epsilon^i{}_j \widetilde\xi_n^j}
\nonumber\\
    &=& 2i\left(\omega_n\int dV \rho|\widetilde\xi|^2
        -2\,\Omega\ \Im\int dV \rho\widetilde\xi_n^{\varpi*}\widetilde\xi_n^{\hat\phi}
        \right)
\nonumber\\
    &=& 2i\omega_n\kappa_n\,\inner{ \widetilde\xi_n}{ \rho\widetilde\xi_n}\,,
\eea
where
\be
 \kappa_n
     = 1 - 2\frac\Omega\omega_n\, \Im\frac{\int dV \rho\widetilde\xi_n^{\varpi*}\widetilde\xi_n^{\hat\phi}}{\int dV\rho|\widetilde\xi_n|^2}\,.
\label{a:kappa_n}\ee

The mode functions $\widetilde\xi_n$ are then given in terms of the $\widehat\xi_n$ of
Eq.~(\ref{a:rhonorm}) by
\be
   \widetilde\xi_n = \frac1{\sqrt{2i\omega_n\kappa_n}}\, \widehat\xi_n\,,
\ee
and we obtain Eq.~(\ref{e:sumxin0}) for the exponentially growing solution prior to
saturation,
\be
\xiII^i = \sum_n \Re\left[ \frac1{i\kappa_n\omega_n (2\beta - i\omega_n)}
        \inner{\widehat\xi_n}{F} \widehat\xi_n^i
      \right]\,.
\label{a:sumxin}\ee
After saturation, the displacement oscillates about an equilibrium position given by
\be
\xiII^i = \sum_n \Re\left[ \frac1{\kappa_n\omega_n^2}
        \inner{\widehat\xi_n}{F} \widehat\xi_n^i
      \right]\,,
\label{a:sumxinf}\ee
where $F$ is the value of the forcing term at saturation.

\end{document}